\documentclass[sigconf, nonacm]{acmart}
\usepackage{popets}

\AtBeginDocument{%
  }

\setcopyright{popets}
\copyrightyear{2023}
\acmYear{2023}
\acmVolume{XXXX}
\acmNumber{Y}
\acmDOI{XXXXXXX.XXXXXXX}
\acmISBN{}
\acmConference{Proceedings on Privacy Enhancing Technologies}
\usepackage[linesnumbered,ruled]{algorithm2e}
\usepackage{amsmath}
\graphicspath{{./figs/}}
\usepackage{subcaption}
\usepackage{multirow}

\makeatletter
\DeclareRobustCommand\onedot{\futurelet\@let@token\@onedot}
\def\@onedot{\ifx\@let@token.\else.\null\fi\xspace}
\def\eg{\emph{e.g}\onedot} 
\def\ie{\emph{i.e}\onedot}

\usepackage{todonotes}
\usepackage{cleveref}
\usepackage{soul}

\newcommand{\syn}{X_\text{syn}}
\newcommand{\ori}{X_\text{ori}}
\newcommand{\train}{X_\text{train}}
\newcommand{\control}{X_\text{control}}
\newcommand{\Nattacks}{N_A}

\newcommand{\GM}{\mathcal{G}} 

\newif\ifblind
\blindfalse

\def\name{\textsf{A\-no\-ny\-me\-ter}\xspace}

\usepackage{ifthen}
\newboolean{review}  
\setboolean{review}{true}
\ifthenelse{\boolean{review}}{

\newcommand{\str}[1]{\st{#1}}
\newcommand{\out}[1][]{\textcolor{red}{[...]}}
}
{

\newcommand{\out}[1][]{\textcolor{blue}{}}
\newcommand{\str}[1]{}
}

\begin{document}

\title{A Unified Framework for Quantifying \\Privacy Risk in Synthetic Data}\titlenote{Accepted at \textit{23rd Privacy Enhancing Technologies Symposium}~(PETS~2023).}


\author{Matteo Giomi}
\email{matteo@statice.ai}
\affiliation{%
  \institution{Statice}
    \city{Berlin}
    \country{Germany}
}
\author{Franziska Boenisch}
\email{franziska.boenisch@vectorinstitute.ai}
\affiliation{%
  \institution{Vector Institute}
    \city{Toronto}
  \country{Canada}
  } 
\author{Christoph Wehmeyer}
\email{chris@statice.ai}
\orcid{0000-0002-9526-0328}
\affiliation{%
  \institution{Statice}
    \city{Berlin}
    \country{Germany}
}
\author{Borbála Tasnádi}
\email{borbala@statice.ai}
\affiliation{%
  \institution{Statice}
    \city{Berlin}
    \country{Germany}
}



\begin{abstract}
Synthetic data is often presented as a method for sharing sensitive information in a privacy-preserving manner by reproducing the global statistical properties of the original data without disclosing sensitive information about any individual. In practice, as with other anonymization methods, synthetic data cannot entirely eliminate privacy risks. These residual privacy risks need instead to be \emph{ex-post} uncovered and assessed. However, quantifying the actual privacy risks of any synthetic dataset is a hard task, given the multitude of facets of data privacy.\\
We present \name, a statistical framework to jointly quantify different types of privacy risks in synthetic tabular datasets. We equip this framework with attack-based evaluations for the \emph{singling out}, \emph{linkability}, and \emph{inference} risks, which are the three key indicators of factual anonymization according to data protection regulations, such as the European General Data Protection Regulation (GDPR). To the best of our knowledge, we are the first to introduce a coherent and legally aligned evaluation of these three privacy risks for synthetic data, as well as to design privacy attacks which model directly the singling out and linkability risks.\\
We demonstrate the effectiveness of our methods by conducting an extensive set of experiments that measure the privacy risks of data with deliberately inserted privacy leakages, and of synthetic data generated with and without differential privacy. Our results highlight that the three privacy risks reported by our framework scale linearly with the amount of privacy leakage in the data. Furthermore, we observe that synthetic data exhibits the lowest vulnerability against linkability, indicating one-to-one relationships between real and synthetic data records are not preserved. Finally, with a quantitative comparison we demonstrate that \name outperforms existing synthetic data privacy evaluation frameworks both in terms of detecting privacy leaks, as well as computation speed. To contribute to a privacy-conscious usage of synthetic data, we publish \name as an open-source library.\footnote{\ifblind URL anonymized for review \url{} \else
\url{https://github.com/statice/anonymeter}\fi}
\end{abstract}



\keywords{anonymization, privacy risks, synthetic data, GDPR, differential privacy}

\maketitle

\section{Introduction}

Societies in the digital era are faced with the challenge of striking a balance between the benefits that can be obtained by freely sharing and analyzing personal data, and the dangers that this practice poses to the privacy of the individuals whose data is concerned. Replacing original and potentially sensitive data with some synthetic data, \ie, data that is artificially generated rather than coming directly from real individuals, is one of the approaches that attempt to resolve this tension~\cite{rubin1993statistical,reiter2002satisfying, bellovin2019privacy}.  
Synthetic data captures population-wide patterns of the underlying potentially sensitive data while ``hiding'' the characteristics of the individuals. Popular approaches for synthetic data generation rely on deep generative models such as Generative Adversarial Networks (GAN)~\cite{goodfellow2014generative} and Variational Autoencoders (VAE)~\cite{VAEs}. These generative models are trained on the original and potentially sensitive data to produce a synthetic dataset that preserves the \emph{utility} of the original data. 

Intuitively, if the generative models are able to generalize well, the synthetic data should not reflect the particular properties of any individual original record. This intuition underpins the use of synthetic data as a privacy-enhancing technology. Unfortunately, assuming that synthetic data simply carry no privacy risks---despite being tempting---is too simplistic. Generative models with enough capacity to express complex data patterns are often found to match the original data too closely~\cite{GeneralizationMemorization, Overlearning, SecretSharer}. 
As a consequence, the synthetic data will likely present some residual privacy risk. Reliably quantifying the privacy risk of synthetic data is therefore an important, yet still not settled, problem---even though such an assessment is not just desirable, but often represents a requirement imposed by legal frameworks, such as the European General Data Protection Regulation (GDPR~\cite{GDPR}), the Canadian Personal Information Protection and Electronic Documents Act (PIPEDA ~\cite{canadaPersonalInformationProtection2019}), an California's Consumer Privacy Act (CCPA~\cite{stateofcaliforniadepartmentofjusticeCaliforniaConsumerPrivacy2018}).

Differential Privacy (DP)~\citep{Dwork.2006Differential} provides a theoretical framework for upper-bounding privacy leakage. However, a gap exists between the worst-case privacy guarantee of DP and what can be empirically measured for practical attacks~\cite{DPAdversaryBounds}. Moreover, due to the stochastic training and inference of generative models, the magnitude of the effective privacy risks exposed by the generated synthetic data cannot be quantified in advance~\cite{groundhog}.  Instead, the residual privacy risks need to be measured a posteriori in an empirical fashion from the generated data. However, since there exist various notions of \emph{practical privacy} (\eg, membership privacy, attribute privacy \cite{PrivacyMetricSurvey}), there is no unified metric for such a measurement. Instead, different metrics have been proposed to measure privacy risks for anonymized data~\cite{PrivacyMetricSurvey}. Yet, interpreting and combining them in a meaningful way is still a topic subject to research.

In this work, we set out to close this gap by proposing \name, an empirical and statistical framework that measures privacy risks in anonymized tabular datasets. \name implements a general three-step procedure for risk assessment based on (1) performing privacy attacks against the dataset under evaluation, (2) measuring the success of such attacks, and (3) quantifying the exposed privacy risk in a well-calibrated and coherent manner. Each of the three steps is connected to the others via common interfaces to keep the framework modular and to allow the same risk quantification method to be shared by different privacy risks.
For each privacy attack, the final risk is obtained by comparing the results of the privacy attack against two baselines; the first resulting from performing the same attack on a control dataset from the same distribution as the dataset under evaluation, and the second resulting from performing a random attack against the dataset under evaluation.
While the latter provides insights into the strength of the main attack, the former makes it possible to measure how much of the attacker's success is simply due to the \emph{utility} of the synthetic data, and how much is instead an indication of \emph{privacy violations}.
Within this framework, we propose three attacks to quantify the risks of \emph{singling out}, \emph{linkability}, and \emph{inference}, respectively---the three privacy metrics defined by the Article 29 Data Protection Working Party~\cite{29WP}, making our framework legally aligned. 

We provide an experimental validation of our framework by testing its ability to detect different amounts of privacy leaks. 
Our results demonstrate that the framework is able to detect these leaks, that the reported risks scale linearly with the amount of privacy leaks present in the dataset, and that risk-computation is efficient even on large datasets. We also show (\Cref{sec:comparison}) that \name outperforms existing evaluation frameworks for synthetic data~\cite{groundhog} in both computational performance and quality of privacy assessment.
Furthermore, in our experiments (\Cref{subsec:synth_eval}), synthetic data exhibits the highest risks to inference and singling out attacks, whereas the risk to linkability is comparably low over all datasets evaluated.
This provides empirical evidence for the common intuition that generating synthetic data breaks the one-to-one links between data records (see, \eg, ~\cite{synth_data_linkage, attacks_on_quasi_identifiers}).
As expected, introducing DP into the training of the generative models~\cite{Abadi} causes a general decrease in the privacy risks reported by \name; besides, a higher utility of the generated data corresponds to a higher reported risk, \ie, the more the synthetic data is close to the original data the higher the risk reported by \name.

In this work, we propose implementations of singling out, linkability, and inference privacy attacks. Yet, \name's modularity allows for a simple and consistent integration of additional attack-based privacy metrics.
\name is designed to be widely \textit{usable} and to provide \textit{interpretable} results, requiring minimal manual configuration and no expert knowledge besides basic data analysis skills. It is also \textit{applicable} to a wide range of datasets and to both numerical and categorical data types. \name is \textit{sensitive} and able to identify and report even small amounts of privacy leaks. Although developed for the specific case of synthetic data, our tool does not make any assumption on how the data is created, except for requiring consistency of attributes and data types, and can also be applied to assess other forms of anonymization and pseudonymization.

The software implementing our privacy evaluations is released as an open-source~library.\footnote{\ifblind URL anonymized for review \url{} \else
\url{https://github.com/statice/anonymeter}\fi}

In summary, we make the following contributions: 
\begin{itemize}
    \item We propose \name, a statistical framework to derive well-calibrated privacy risks against different types of privacy attacks in a coherent way. 
    \item To the best of our knowledge, we are the first to introduce a practical tool for jointly evaluating all the three privacy risks (singling out, linkability, and inference) defined by the main privacy regulations~\cite{29WP}.
    \item For each risk, we provide a concrete attack algorithm, its implementation, and thorough experimental evaluation. Our framework compares favourably to previous work on privacy evaluation of synthetic datasets (see \Cref{sec:comparison}).
    \item We are the first to propose a practical implementation for the risk of singling out, and a linkability attack directly modeling the possibility of linking records belonging to specific individuals, rather than relying on related concepts, such as membership inference~\cite{groundhog}.
    \item We provide our framework, including the attack implementations, as an open-source library to support practical privacy-enhanced synthetic data usage.
\end{itemize}
\section{Notation and Background}\label{sec:background}

The following notation is used in the rest of the paper: a tabular dataset $X$ is a collection of $N$ records $\mathbf{x}=(x_1, ..., x_d)$, each with $d$ attributes, drawn from a distribution $\mathcal{D}$. We use the subscripts $ori$ and $syn$ to denote original datasets, \ie, collections of data records sampled from $\mathcal{D}$, and synthetically created datasets, respectively. More in detail, we denote an original dataset by $\ori = \{\mathbf{x}_\text{ori}^{1}, ..., \mathbf{x}_\text{ori}^{N}\}$ and a synthetic dataset by $\syn = \{\mathbf{x}_\text{syn}^{1}, ..., \mathbf{x}_\text{syn}^{M}\}$. We make use of matrix notation to indicate columns in the datasets: $X[:, i]$ is a vector of size $N$ containing the $i\textsuperscript{th}$ attribute of all the records, and $\mathbf{x}[i] = x_i$ is the value of the $i\textsuperscript{th}$ attribute of record $\mathbf{x}$. Finally, $\GM$ denotes the generative model from which the synthetic dataset $\syn$ is produced.

\subsection{Synthetic Data Generation}

In general, synthetic data is produced by a generative model $\GM$ that is supposed to learn the distribution $\mathcal{D}$. However, since this distribution is usually unknown, the model $\GM$ is instead trained on $\ori$ sampled from $\mathcal{D}$. Once trained, the model $\GM(\ori)$ can be understood as a stochastic function that, without any input, generates synthetic data records $\mathbf{x}_\text{syn}$~\cite{groundhog}. By querying $\GM$ multiple times, a full synthetic dataset $\syn$ can be sampled. Ideally, the generated synthetic data should reflect most of the statistical properties of the distribution $\mathcal{D}$. Yet, since $\GM$ only has access to $\ori$, and only learns a partial representation of the data distribution, the generated data can only approximate $\mathcal{D}$.

Several methods exist to generate synthetic data. A possibility is to use statistical models, such as Bayesian networks~\cite{young2009using} or Hidden Markov models~\cite{dainotti2008internet, ngoko2014synthetic}. Such models generate explicit parametric representations of $\mathcal{D}$ and the features to be extracted from $\ori$ are determined beforehand. In contrast, deep learning models for synthetic data generation, such as GANs~\cite{goodfellow2014generative} and VAEs~\cite{VAEs}, learn which attributes to extract during a stochastic training process.

\subsection{Privacy-Preserving Synthetic Data}

One main reason to generate synthetic data is for the purpose of privacy-preserving data releases and data sharing: synthetic datasets are supposed to reproduce properties of an original dataset~$\ori$ from $\mathcal{D}$ without containing the personal data from ~$\ori$~\cite{abay2018privacy, groundhog}. Yet, recent studies indicate that through high-utility synthetic data it is still possible for an attacker to extract sensitive information about the original data~\cite{groundhog}.

In this work, we use CTGAN~\cite{xu2019modeling} to generate the synthetic data for our experiments. 
CTGAN is a GAN that uses a conditional generator to enable the generation of synthetic tabular data with both discrete and continuous-valued columns. The approach uses a mode-specific normalization as an improvement for non-Gaussian and multimodal data distributions~\cite{xu2019modeling}.
Privacy guarantees can be integrated into CTGAN using DP~\cite{Dwork.2006Differential}. In this case, the generative model is trained with a DP optimizer~\cite{Abadi}. As a result of the post-processing robustness of DP, the synthetic data generated from such DP models also enjoy the same level of privacy guarantee as the trained generative model~\cite{zhang2017privbayes,jordon2018pate, qu2019gan, abay2018privacy}.

\subsection{Privacy Metrics and Attacks in Synthetic Data}

Privacy is a multi-faceted concept, which is reflected in the availability of dozens of different privacy metrics---see, e.g.,~\cite{PrivacyMetricSurvey} for an overview. In the concrete case of measuring the privacy leakage of synthetic data, many studies rely on similarity tests~\cite{yale2019assessing}, on distance metrics calculating the mean absolute error between original and generated data records~\cite{mendelevitch2021fidelity}, or on measuring the number of identical records between original and synthetic datasets~\cite{Lu2019Empirical}. When the synthetic data is generated with DP guarantees, the DP privacy budget, usually denoted by~$\varepsilon$, can also be used to report on the privacy of a synthetic dataset. However, for most of these metrics, it is unclear how they translate into privacy implications in practice and what concrete privacy risks exist for individual data records~\cite{DPAdversaryBounds}. Therefore, using the success rate of concrete privacy attacks is becoming a common approach to quantifying the privacy of synthetic data, \eg~\cite{groundhog}.
In this work, we integrate into \name the evaluation of \textit{the three} privacy attacks that anonymization techniques must protect from, according to the GDPR~\cite{29WP}: \emph{singling out}, \emph{linkability}, and~\emph{inference}.
We close the gap of prior evaluation frameworks~\eg, \cite{groundhog} that only jointly consider a subset of these legally essential risks.
The importance to consider these three risks against anonymization results from their implication on individuals' privacy:
singling out can be seen as a way to indirectly identify a person in a dataset. At the same time, it can serve as a stepping stone towards linkage attacks~\cite{cohen2020towards}, which have been shown to yield complete de-anonymization of datasets in the past~\cite{sweeney2000simple,narayanan2008robust}. Inference attacks, in turn, can disclose highly sensitive information on individuals, such as their genomics~\cite{ayday2017inference}.

\textbf{Singling out} happens whenever it is possible to deduce that within the original dataset there is a single data record with a unique combination of one or more given attributes~\cite{29WP}. For example, an attacker might conclude that in a given dataset $\ori$ there is exactly one individual with the attributes gender: \emph{male}, age: \emph{65}, ZIP-code: \emph{30305}, number of heart attacks: \emph{4}.
It is important to note that singling out does not imply re-identification. Yet the ability to isolate an individual is often enough to exert control on that individual, or to mount other privacy attacks~\cite{boenisch2021side}.

\textbf{Linkability} is the possibility of linking together two or more records (either in the same dataset or in different ones) belonging to the same individual or group of individuals~\cite{29WP}. This can be used for de-anonymization, as proven in the past~\cite{Netflix, sweeney2002k}. 
Due to statistical similarities between the generated data and the original data, linkability risks may still exist in synthetic datasets.

\textbf{Inference} happens when an attacker can confidently guess (infer) the value of an unknown attribute of the original data record~\cite{29WP}. An example of successful inference would consist in the attacker being able to confidently deduce that a record in the original dataset $\ori$, with attributes ``gender'': \emph{male}, ``age'': \emph{65}, ``ZIP-code'': \emph{30305} holds the secret attribute ``number of heart attacks'': \emph{4}. 

When measuring privacy risks, an important distinction has to be made between what an attacker can learn at a population level (generic information) and on an individual level (specific information). Generic information is what provides utility to the anonymized data; specific information enables the attacker to breach the privacy of some individuals~\cite{BlueBook}. 
\name distinguishes what the attacker learns from the anonymized dataset as generic information from specific inference, thus quantifying the privacy risk (see \Cref{subsec:architecture}).
\section{Related work}
\label{sec:related_work}
In this section, we present related work on privacy attacks and privacy evaluation of (synthetic) datasets.

\textbf{Privacy Attacks:} An extensive theoretical analysis and formalization of \emph{singling out} has been carried out in~\cite{PredicateSinglingOut}. Such analysis relies on the knowledge of the population distribution $\mathcal{D}$, which requires complex and careful modeling, \eg \cite{RocherUniqueness}. That would make the evaluation harder to use and to interpret. We will instead approximate the baseline probability by carefully comparing the success of the attack on different samples from the data. To the best of our knowledge, we are the first to propose a singling out attack algorithm for tabular synthetic data, and to implement it in a practical and usable framework.

The concepts of \emph{linkability} and \emph{inference} attacks have extensively been studied both on synthetically generated datasets~\cite{groundhog} and datasets that were anonymized with other approaches~\cite{Netflix, sweeney2002k, Herranz2015Revisiting}. While many practical inference attacks exist, \eg ~\cite{Netflix, groundhog}, our work is the first one to devise a methodology to measure directly the linkability risk posed by the release of tabular "anonymized" data, and to implement it in a practical and efficient algorithm.
In addition to empirical demonstration of the power of those attacks, theoretical analyses have been introduced, which formalize those concepts and show that the success rate of an attack can be improved when an attacker holds auxiliary information about rare attributes~\cite{Merener2012Theoretical}.
Finally, \emph{background knowledge attacks} have been proposed, in which an attacker holds knowledge about target individuals and uses it to reconstruct their information~\cite{olatunji2021review} and to perform \emph{membership inference}~\cite{li2013membership}, \ie, deduce the presence or absence of a particular individual in the original dataset. 

\textbf{Privacy Evaluation Frameworks:} 
Similar to \name, other frameworks exist that rely on using privacy metrics and attacks to evaluate the privacy of synthetic datasets.
Yet, to the best of our knowledge, we are the first in this line of work to leverage a control dataset for estimating specific privacy risk in contrast to generic population-wide risk.
\cite{Lu2019Empirical} uses similarity metrics between original and generated data records to quantify the risk of re-identification for data generated with GANs. \cite{Kuppa2021Towards} calculates a per-data record privacy by measuring a memorization coefficient based on distance measures between original and synthetic data records in a latent space.
\cite{Acs2016Testing} introduces the framework of differential testing, similar to the notion of DP, which deems a dataset anonymized when the inference accuracy from the synthetic data is about the same whether the user’s record is included in the original dataset or not.
In the medical domain, \cite{Oprisanu2021Utility} and \cite{mendelevitch2021fidelity, ElEmam} evaluate privacy vs. utility of generative models and synthetic datasets, respectively. The first two works rely on the notion of membership inference to quantify the privacy risks, the last one on equivalence classes and intra-record distances.
Finally, attack-based approaches have been used to quantify the strength of database query anonymization mechanisms~\cite{aircloak}.

The framework for privacy evaluation in synthetic data that is most closely related to ours is~\cite{groundhog}. 
It operates in the same scenario as we do by treating the data generation mechanism as a black box and by evaluating the privacy leakage directly on the generated synthetic data.
The privacy leakage is measured in terms of a \emph{privacy gain} ($PG$), which is defined as the difference between the \emph{advantage} of an adversary trying to guess secrets about a target data record given the original data compared to an adversary with access only to the synthetic data.
The attack is formalized as an adversarial game.
For a given target, such games are repeated many times, each time requiring training of a new generative model and sampling several synthetic datasets. As a consequence, the computational cost of the attack is very high which severely limits its usability. As detailed in \Cref{sec:comparison}, even without taking the training of the generative models into account, \cite{groundhog}'s methodology is ten times slower than~\name.

Additionally, our privacy risk metric has several advantages over the Privacy Gain defined in ~\cite{groundhog}. First, it is closer to the legal terminology and therefore easier to understand and explain than $PG$. Second, contrary to our work, in \cite{groundhog} there is no separation between how much of the attacker's advantage is due to actual privacy violations and how much of it is instead due to the utility of the synthetic data. Third, $PG$ is measured from the evaluation of a few records, typically hand-picked outliers. It is not clear how a characterization of the privacy of a restricted sample of records can be extrapolated to a whole dataset. The privacy risk computed by \name is instead obtained by attacking thousands of different targets, making it a more robust metric compared to $PG$. \name also naively provides an estimate of the statistical uncertainties on the reported risk. Moreover, as we will see in \Cref{subsec:architecture}, \name incorporates sanity checks ensuring that the risk estimate is based on attack models that are actually effective.

Finally, \cite{groundhog} does not include any evaluation of the singling out risk. Also, in~\cite{groundhog} the linkability $PG$ is not measured directly, but instead is indirectly derived from a membership inference game.\footnote{However, linkability cannot be fully measured through membership inference. This is because linkability is a stronger privacy breach. In fact, establishing a successful linkage between a target $t$ and some records in a dataset $X$ implies membership of $t$ in $X$. Yet, the opposite is not true, \ie, even if membership of $t$ in $X$ is established, the attacker still does not know which records of $X$ are directly linked to the target.}

\section{\name Framework}

We propose a framework which provides a coherent assessment of diverse privacy risks based on dedicated privacy attacks.

\subsection{Threat Model and Assumptions}\label{subsec:assumptions}

To provide a conservative privacy risk assessment, we consider the strongest threat model in which the attacker is in full possession of the synthetic dataset.
Moreover, the attacker holds additional partial but correct knowledge, called \emph{auxiliary information}, about a subset of the original records (the \emph{target} records). This accounts for practical scenarios where overlapping data sources are common.
Depending on the amount and quality of the auxiliary information, more or less powerful attacks can be modeled. 
We rely on simple yet effective heuristics to choose the auxiliary knowledge for our respective attacks (see~\Cref{sec:attacks} for details).
The targeted original records are chosen at random from the original dataset~$\ori$. That is, we make no assumption on how the attacker would choose the targets, resulting in a more robust evaluation of the overall privacy offered by the synthetic data. If needed, however, the framework can easily be adapted to attack specific records, for example to measure privacy risks for some specific sub-population in the data, or particular individuals.

We treat the data generation mechanism as a black box that \textit{cannot} be accessed or queried by the attacker who only receives the synthetic dataset. Concerning the original dataset $\ori$, we assume that it consists of $N$ records drawn independently from the population $\mathcal{D}$, and that each record refers to a different individual. The full original dataset is split into two disjoint sets $\train$ and $\control$. That is, $\ori = \train \cup \control$ and $\train \cap \control = \emptyset$. 
The synthetic dataset $\syn$ is sampled from a generative model trained on $\train$ exclusively: $\syn \sim \mathcal{G}(\train)$. 
In \Cref{app:control_ablation} we show that, in the scope of \name where $\control$ consists of at least a few thousand data records, training the generative model on $\train$ instead of on the full $\ori$ does not affect the utility of the resulting synthetic data.
To evaluate privacy, \name requires all three datasets, $\train$, $\syn$, and $\control$. All the datasets have the same number of attributes $d$, but the number of records might differ.

\subsection{Framework Architecture}\label{subsec:architecture}

\begin{figure}[t]
\centering
\includegraphics[trim=0 1cm 0cm 0cm,clip, width=0.45\textwidth, page=3]{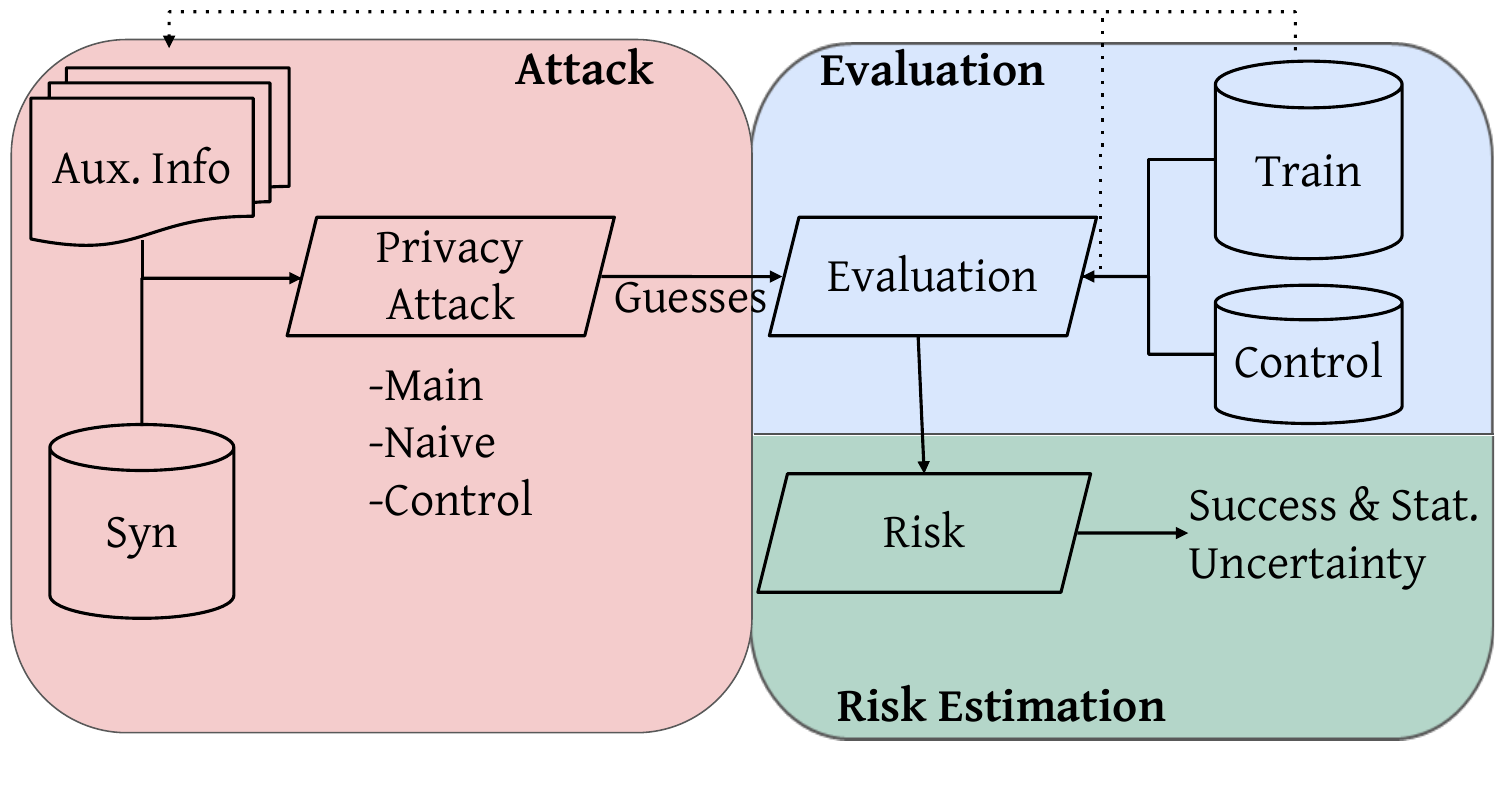}
\caption{Schematic overview of our framework. The attacker is given access to the entire generated synthetic dataset and some auxiliary information. 
(1) In the \emph{attack phase}, the framework performs three different attacks (main, naive, and control) each of which outputs guesses on the original private data.
(2) The correctness of these guesses is then evaluated in the \emph{evaluation phase} against the original training data (main and naive attack) or the control data (control attack).
(3) Based on the evaluation, a final statistical risk quantification is output in the \emph{risk estimation} phase.}\label{fig:main_schema}
\end{figure}

Privacy risks in our framework are estimated following a common procedure, see \Cref{fig:main_schema}. In this modular design, the analysis is made up of three sequential steps: (i) the \emph{attack phase} in which the privacy attacks are carried out, (ii) the \emph{evaluation} phase in which the success of the attacks is measured, and (iii) the \emph{statistical risk estimation} phase quantifying the privacy leakage for the attack. How the concrete attacks and evaluation phases are carried out differs for each evaluated privacy risk, but the way the risks are derived from the results of the evaluation phase is common to all cases. This consistent view improves the interpretability of the different attack results and makes \name more modular. 

\textbf{Attack Phase:} the attack phase consists of executing three different attacks.
First, the ``main'' privacy attack in which the attacker uses the synthetic dataset $\syn$ to deduce private information of records in the training set $X_\text{train}$. Second, a ``naive'' attack is carried out based on random guessing, to provide a baseline against which the strength of the ``main'' attack can be compared. Finally, to distinguish the concrete privacy risks of the original data records (\ie, specific information) from general risks intrinsic to the whole population (\ie, generic information), a third ``control'' attack is conducted on a set of control records from $\control$.
For all the risks, each of the three attacks is formulated as the task of making a set of guesses: $\textbf{g} = \{g_1, ..., g_{\Nattacks}\}$ on $\Nattacks$ original \emph{target} records. As an example, a singling out guess could state that “there is just one person in the original dataset who is male, 65 years old and lives in area 30305”.
The naive attack draws its guesses at random, using the synthetic dataset only to know the domain of the dataset attributes. The main and the control attacks generate the guesses trying to actively leverage the synthetic dataset (and the auxiliary information, when available) to gain information. They both share the same attack algorithm, but in the main privacy attack such guesses are evaluated against $\train$, whereas in the control attack the guesses are evaluated against $\control$. Note that $\control$ is completely independent of the synthetic data generated from $\train$ (see \Cref{subsec:assumptions}). 
Hence, if the attacker is successful in guessing information about records in $\control$, this must only be due to patterns and correlations that are common to the whole population $\ori$, rather than being specific to some training record. Therefore, the difference between the success rate of the two attacks can provide a measure of privacy leakage that occurred by training $\mathcal{G}$ on $\train$:
\name reports a privacy leakage when the attacker is more successful at targeting $\train$ than $\control$.

\textbf{Evaluation Phase:} In the evaluation phase, the guesses from the attack phase are compared against the truth in the original data to estimate the privacy risk. The outcome of the evaluation phase is a vector of bits $\mathbf{o} = \{o_1, ..., o_{\Nattacks}\}$, where $o_i=1$ if the $i\textsuperscript{th}$ guess $g_i$ is correct, otherwise $o_i = 0$. Each attack defines the criteria for a guess to be considered correct. In the singling out example from above, the guess would be considered correct if there indeed exists exactly one such individual in the original data.

\textbf{Risk Quantification Phase:} In the risk quantification phase success rates of the ``main'' privacy attack are derived from the evaluation together with a measure of the statistical uncertainties due to the finite number of targets.
Under the assumption that the outcome $o_i$ of each attack is independent from the others, $\mathbf{o}$ can be modeled as Bernoulli trials. We define the true privacy risk $\hat{r}$ as the probability of success of the attacker in these trials. The best estimate $r$ of the true attacker success rate $\hat{r}$ and the accompanying confidence interval $\hat{r} \in r \pm \delta_r$ for confidence level $\alpha$ are estimated via the Wilson Score Interval~\cite{WilsonInterval}:
\begin{equation}\label{eq:wilson}
\begin{gathered}
    r = \frac{N_S + z_{\alpha}^2/2}{\Nattacks + z_{\alpha}^2} \\
    \delta_r = \frac{z_{\alpha}}{\Nattacks + z_{\alpha}^2}\sqrt{\frac{N_S (\Nattacks-N_S)}{\Nattacks} + \frac{z_{\alpha}^2}{4}}\text{,}
\end{gathered}
\end{equation}
with $N_S = \sum_{i=1}^{\Nattacks} o_i$ being the total number of correct guesses, and $z_{\alpha}$ the probit, \ie, the inverse of the cumulative distribution function of the normal distribution, corresponding to the confidence level $\alpha$. 
Using \Cref{eq:wilson} and the number of successful guesses for each attack, we evaluate the success rates for the ``main'', ``naive'', and ``control`` attacks as $(r_\text{train} \pm \delta_\text{train})$, $(r_\text{naive} \pm \delta_\text{naive})$, and $(r_\text{control} \pm \delta_\text{control})$, respectively.

The success rate of the naive attack provides a baseline to measure the strength $s$ of the attack which can be defined as the difference between the success rate of the main attack against training records and the success rate of the naive attack:

\begin{equation}\label{eqn:strenght}
    s = r_\text{train} - r_\text{naive}
\end{equation}

with the error on $s$ obtained via error propagation as $\delta_s = \sqrt{\delta_r^2 + \delta_{naive}^2}$. 

If the attack is weaker than the naive baseline, \ie, $r_\text{naive} \geq r$, the attack is said to have failed. This can happen in case of incorrect modeling, for instance when the attacker is given too little auxiliary information or auxiliary information that is uncorrelated with the guess's objective, or when the synthetic dataset has little utility and it is actually misleading for the attack. In such cases, \name warns the user that the results are considered void of meaning and must be excluded from the analysis. See \Cref{tab:fraction_successful_attacks} in \Cref{subsec:synth_eval} for details on the fractions of generated valid attacks for our three concrete instantiations of singling out, linkability and inference risks. Note that these fractions depend on the concrete attacks and their respective instantiations. Excluding invalid attacks is important in practice, because it avoids the situation in which ``no risk'' is reported due to the incorrect modeling of the attacks.

For the ``control'' attack, the attack's success rate is evaluated on control records ($r_\text{control}$) using \Cref{eq:wilson}. Intuitively, if the synthetic dataset contains more information on the training records than on those in the control set, this implies $r_\text{train} \geq r_\text{control}$. From these two success rates the specific privacy risk $R$ is derived as as:

\begin{equation}\label{eq:risk}
   R = \frac{r_\text{train}-r_\text{control}}{1-r_\text{control}}. 
\end{equation}

The numerator in the above expression corresponds to the excess of attacker success when targeting records from $\train$ versus the success when the targets come from $\control$. The denominator represents the maximum improvement over the control attack that a perfect attacker ($r=1$) can obtain, and helps contextualizing the difference at the numerator by acting as a normalization factor. For example, suppose that out of 100 guesses the attacks against the training and control sets are correct 90 and 80 times, respectively: $r_\text{train} = 0.9$ and $r_\text{control}=0.8$. Of the 90 correct guesses of the main privacy attack, 80 could be explained as being due the utility of the dataset, leaving the remaining 10 correct guesses to indicate privacy violations. This $0.1$ excess in the success rate $r_\text{train}$ translates in a $R=0.5$ risk, since the best possible attack can only score 100 out of 100 guesses, \ie, its rate can only be $0.2$ higher than $r_\text{control}$. Similar ways of normalizing the risk are proposed in~\cite{global_privacy_risk, aircloak}, with the difference that in our work the normalizing baseline ($r_\text{control}$) is derived from attacking a control set of records, rather than from the success of the naive attack.

If both success rates are identical, access to the synthetic data does not give the attacker any benefit to gain information about the training data: the success of the attack can be explained by the general utility of the synthetic data, \ie, it is a consequence of general inference. If, however, the success rate on training data exceeds the one on control data, this shows that information has been leaked from the synthetic data. 

\textbf{Properties of our privacy risk:}
There are no unified requirements on properties that privacy metrics should possess~\cite{PrivacyMetricSurvey}.
Our privacy risks align best with the three main requirements of the privacy metrics formulated by~\cite{shokri2011quantifying}.
First, we evaluate the \emph{correctness} of the guesses generated by the attacker; second, we report the \emph{uncertainty} of the risks through confidence intervals; and third, as demonstrated empirically in Sections~\ref{subsec:linearity} and \ref{sec:comparison}, our proposed metric is accurate as it is able to measure the actual percentage of data leaked from the synthetic dataset.
The experiments of Sections~\ref{subsec:linearity} and \ref{sec:comparison} also show that our metric is \emph{meaningful} in the sense that $R\sim0$ if (and only if) the evaluated dataset is independent of the original data (noninterference), and that the reported risk increases proportionally with the amount of privacy leaks~\cite{smithQIF}.
Finally, our privacy risks are based on probabilities, namely the probabilities of making correct guesses on the sensitive data, as suggested for privacy metrics by~\cite{andersson2007fundamentals}.

\subsection{Practical Privacy Evaluation Bounds}

In general, an attack-based privacy analysis provides a \emph{lower} bound for the privacy risk (in contrast to theoretical frameworks, such as DP~\cite{Dwork.2006Differential} that provide \emph{upper} bounds, \ie, worst-case guarantees). 
Therefore, the computed privacy risks are just as representative as the employed attacks are.
Yet, in practice, using attack-based approaches to quantify privacy leakage has become state-of-the-art in several domains, such as machine learning, \eg~\cite{murakonda2020ml,nasr2021adversary}.

To overcome potential limitations of an attack-based approach, we model powerful and knowledgeable attackers: we always assume the attacker holds knowledge of the entire synthetic dataset---the worst case scenario in which the synthetic data is released to the public. Note that in many practical applications, the value of the datasets which are processed, see \eg \cite{gartner_syn_data, gartner_report,  forbes_syn_data, tech_crunch, gartner_predictions}, discourages such scenarios. In addition, for the linkability and inference estimation, partial but correct auxiliary knowledge of some original records is also available to the attacker. Finally, we evaluate the attack strength by comparing to a baseline attack based on the uninformed guesses from the naive attack. This adds the context needed to interpret the results correctly. In particular, the results are only valid if the ``main'' attack is able to outperform the baseline scenario.

\section{Our Framework's Privacy Attacks}\label{sec:attacks}

This section proposes concrete instantiations of three privacy attacks to assess the fundamental risks of \emph{singling out}, \emph{linkability}, and \emph{inference} within our framework. We chose to implement attacks measuring these specific three risks due to their importance in the legislation: according to the GDPR \cite{29WP}, any successful anonymization technique must provide protection against such risks. For each privacy risk, we present the design and implementation of both the attack and the evaluation phases. The risk quantification phase is common to all attacks.

\subsection{Singling out}\label{subsec:singling_out}

The singling out attack is given the task to create $\Nattacks$ many predicates based on the synthetic data that might single out individual data records in the training dataset. As stated above, it produces guesses like: \textit{“there is just one person in the original dataset who is male, 65 years old and lives at area 30305”}. The intuition behind this approach is that attributes (or combinations thereof) that are rare or unique in the synthetic data might also be rare or unique in the original data. Therefore, access to the synthetic data would allow for generating more meaningful predicates than uninformed~guessing.

\textbf{Attack Phase:} For the attack phase, we introduce two algorithms, namely the univariate \texttt{PredicateFromAttribute} and the multivariate \texttt{MultivariatePredicate}, that can be used to generate the $\Nattacks$ many singling-out predicates (\textit{guesses}). While the univariate algorithm creates predicates using single attributes, the multivariate algorithm relies on the combination of several attributes. 

\begin{algorithm}
\SetKwData{predicates}{predicates}

\SetKwFunction{Set}{Set} 
\SetKwFunction{In}{In}
\SetKwFunction{Is}{Is}
\SetKwFunction{Sum}{Sum}
\SetKwFunction{IsContinuous}{IsContinuous} 

\SetKwFunction{PredicateFromAttribute}{PredicateFromAttribute}
\SetKwProg{Fn}{def}{:}{}

\Fn{\PredicateFromAttribute{$X$, $a$}}
    {
    \predicates $\leftarrow$ []\;
    \If{\Sum$(\syn[:, a] == \text{NaN}) == 1$}
        {
        \predicates $+=$ "$a$ \Is $\text{NaN}$"\;
        }
    \If{\IsContinuous$(a)$}
        {
        \predicates $+=$ "$a$ $\leq \min(X_{syn}[:, a])$"\;
        \predicates $+=$ "$a$ $\geq \max(X_{syn}[:, a])$"\;
        }
    {
    \For{$v$ \In \Set$(\syn[:, a])$}{
        \If {\Sum$(\syn[:, a] == v) == 1$}
        {
            \predicates $+=$ "$a$ == v"\;
        }
    }
    }
\Return{\predicates}\;
}

\caption{Creating a univariate singling out predicates for attribute $a$.}
\label{algo:univariate_SO_base}
\end{algorithm}

\Cref{algo:univariate_SO_base} samples all unique attribute values in the synthetic dataset as predicates. For categorical attributes or in case of missing values, such unique values are values that appear only once in the dataset. For numerical continuous attributes, we use the maximum and minimum value of the respective attribute and create the predicate based on being smaller than the minimum or larger than the maximum value. The intuition behind this approach is to exploit outlier values in all the one-way marginals. Such univariate predicates are specially designed to exploit privacy leaks in pre- and post-processing, \eg, when numerical values sampled from the generative models are scaled to ranges derived from the original dataset, or when high-cardinality categories such as identifiers or addresses are preserved. By running \Cref{algo:univariate_SO_base} for all attributes in a dataset, we obtain a large collection of univariate singling-out predicates. The attacker picks $\Nattacks$ of them at random to use them as guesses.

\begin{algorithm}
\SetKwInOut{Input}{input}
\SetKwInOut{Output}{output}
\SetKwData{attributes}{attributes}
\SetKwData{guesses}{guesses}
\SetKwData{predicates}{predicates}
\SetKwData{predicate}{predicate}

\SetKwFunction{Query}{Query} 
\SetKwFunction{Len}{Len}

\SetKwFunction{In}{In}
\SetKwFunction{Is}{Is}
\SetKwFunction{median}{median}
\SetKwFunction{IsContinuous}{IsContinuous} 
\SetKwFunction{LogicalAnd}{LogicalAnd} 

\SetKwFunction{MultivariatePredicate}{MultivariatePredicate}
\SetKwProg{Fn}{def}{:}{}

\Fn{\MultivariatePredicate{$X$, \attributes, $\mathbf{x}$}}{

    \predicates $\leftarrow$ []\;
    \For{$a$ \In \attributes}
            {
            p $\leftarrow$ ""\;
            \If{$(\mathbf{x}[a] == \text{NaN})$}
                {
                p $\leftarrow$ "$a$ \Is $\text{NaN}$"\;
                }
            \eIf{\IsContinuous$(a)$}
                {
                    \eIf{$\mathbf{x}[a] \geq $ \median$(X[:, a])$}
                        {
                        p $\leftarrow$ "$a \geq \mathbf{x}[a]$"\;
                        }
                        {
                        p $\leftarrow$ "$a \leq \mathbf{x}[a]$"\;
                        }
                }
                {
                p $\leftarrow$ "$a == \mathbf{x}[a]$"\;
                }
            \predicates += p\;
            }
\BlankLine
\predicate $\leftarrow$ \LogicalAnd(\predicates)\;
\Return{\predicate}       
}

\caption{Create a multivariate predicate from the attributes of record $\mathbf{x}$.}
\label{algo:multivariate_SO_base}
\end{algorithm}

\Cref{algo:multivariate_SO_base} creates predicates as the logical combinations of univariate predicates created from randomly selected synthetic data records. It starts by drawing a random record $\Tilde{\mathbf{x}}$ from the synthetic dataset and considering a random set of attributes $\{a_1, \ldots, a_d\}$. A multivariate predicate is then formulated as the logical AND of the univariate expressions derived from the values of $\Tilde{\mathbf{x}}$.\footnote{For example, a multivariate singling-out predicate using five attributes and generated from the Adults dataset looks like this: ``race=`Asian-Pac-Islander' \& relationship=`Husband' \& education=`Bachelors' \& capitalloss$\ge$1902 \& country=`Philippines'{''}.}  If attribute $a_i$ is categorical or not a number, the expression sets $a_i$ to be equal to $\Tilde{\mathbf{x}}[a_i]$. If $a_i$ is numerical, the expression sets for values of $a_i$ either greater or equal or smaller or equal than $\Tilde{\mathbf{x}}[a_i]$. The sign of the inequality depends on whether $\Tilde{\mathbf{x}}[a_i]$ is above or below the median of $X_\text{syn}[a_i]$, respectively. This latter condition helps creating predicates with a higher chance of singling out. The attacker evaluates each of these predicates on the synthetic dataset and adds them to the set of guesses only if they are satisfied by a single record in $\syn$. 
The fraction of generated predicates by the multivariate algorithm that passes this selection depends on the dataset and the number of attributes used to generate the guesses. 
For the experiments and datasets presented in \Cref{subsec:synth_eval} this fraction is globally $\sim24\%$, \ie, to obtain $\Nattacks$ singling-out predicates, roughly $4\Nattacks$ predicates must be generated.

We note that starting from randomly selected synthetic records and attributes ensures that the attack predicates explore the whole parameter space while not overfitting to the synthetic dataset.

To quantify the strength of the attack, we implement an algorithm that generates random predicates in order to measure the probability of creating predicates that single out an individual by chance. Such predicates are created as the joined logical \texttt{AND} of univariate predicates of the form: $a~\Pi~v $ where $a$ is a randomly chosen attribute, $\Pi$ a comparison operator\footnote{$=, \neq, >, <, \geq, 
\leq$.} selected at random, and $v$ is a value sampled uniformly from the support of $\syn[:, a]$.

The randomly generated predicates from this algorithm are not evaluated on the synthetic dataset, and reach the evaluation phase of the analysis without undergoing the selection phase.
An experimental evaluation of our attack strength as a function of the number of attributes is presented in \Cref{fig:so_strength} in the Appendix.
It illustrates the effect of the random predicates on the reported risk.

\textbf{Evaluation Phase:} For the univariate and multivariate algorithms, as well as for the naive attack, the results are sets of $\Nattacks$ predicates. These singling out guesses are evaluated on the original dataset to check whether they represent singling-out predicates in the original data as well.

\textbf{Risk Quantification Phase:} As for each of the three privacy attacks, the output of the evaluation is used for risk quantification. To derive a unique singling out risk estimate, both the univariate and multivariate attack algorithms are run, and the one with the best performance (highest risk) is chosen to provide a more conservative privacy assessment.

In contrast to the other privacy attacks, in the case of the singling out attack, care must be taken when comparing the results of the attack against the training set $r_\text{train}$ and the control set $r_\text{control}$. The ability of the attack to single out a record is strongly dependent on the size of the dataset. If, as it is often the case in practice, the control dataset is smaller than the training set, the number of predicates that successfully single out in the control dataset is lower by construction than in the case of singling out in the training set. To be able to measure the true privacy risk with \Cref{eq:risk} we need to know how many predicates would have singled out in a population of size $N_\text{train}$, given the number of predicates that single out in a population of size $N_\text{control}$ (where $N_\text{control} \leq N_\text{train}$). We do this by developing a model based on the Bernoulli distribution which is then fitted to the data to derive the scaling factor needed to compare $r_\text{train}$ and $r_\text{control}$ accounting for the different sample sizes. This model is presented in \Cref{app:so_correction}.

\subsection{Linkability}\label{subsec:linkability}
The linkability attack tries to solve the following task: \textit{"Given two disjoint sets of original attributes, use the synthetic dataset to determine whether or not they belong to the same individual"}.
We assume that there exist two (or more) external datasets A and B containing some of the attributes of a set of original data records and that these attributes are also present in the synthetic data. An example of such a scenario is depicted in \Cref{tab:linkability}. 

\begin{table}[t]
\small
\begin{minipage}{0.48\textwidth}
\centering
        \begin{tabular}{ccccc}
            \multicolumn{5}{c}{Synthetic data} \\
            \toprule
            Gender &  Age & ZIP Code & Heart Attacks & ...\\
            ... &  ...         &  ...       & ...    & ...\\
            ... &  ...         &  ...       & ...    & ...\\
            \bottomrule
        \end{tabular}
\end{minipage}\vspace{0.2cm}
\begin{minipage}{0.24\textwidth}
\centering
        \begin{tabular}{ccc}
            \multicolumn{3}{c}{External dataset $A$} \\
            \toprule
            ... & Gender &  ZIP Code \\
            ... & ... &  ...         \\
            ... & ... &  ...         \\
            \bottomrule
        \end{tabular}
\end{minipage}~
\begin{minipage}{0.24\textwidth}
\centering
        \begin{tabular}{ccc}
            \multicolumn{3}{c}{External dataset $B$} \\
            \toprule
            ... & Age & Heart Attacks \\
            ... & ...        &  ...   \\
            ... & ...        &  ...   \\
            \bottomrule
        \end{tabular}
\end{minipage}
\caption{Setting of the linkability risk evaluation. The original external datasets $A$ and $B$ (bottom, left and right tables, respectively) share some attributes with the synthetic data (top) which can then be used to establish a link between datasets $A$ and $B$.\label{tab:linkability}}
\end{table}
\setlength{\textfloatsep}{-0.1cm}
\textbf{Attack Phase:} 
In the linkability attack, the target records of the attack are a collection $T$ of $\Nattacks$ original records randomly drawn from $X_{ori}$. We assume that the attacker has some knowledge on the targets, \ie, the values of the attributes in datasets $A$ and $B$: $T[:, A]$ and $T[:, B]$. The goal of the attack is then to correctly match records of $T[:, B]$ to each record in $T[:, A]$, or vice versa.

To do so, for every record in $T[:, A]$ the attacker finds the $k$ closest synthetic records in $X_\text{syn}[:, A]$. The resulting indices are $\mathbf{l}^A = (\mathbf{l}^A_i, ..., \mathbf{l}^A_{\Nattacks})$, where each $\mathbf{l}^A_i$ is the set of indexes of the $k$ synthetic records that are nearest neighbours of the $i\textsuperscript{th}$ target in the subspace of feature set $A$. The same procedure is repeated on the feature set $B$, resulting in the indexes $\mathbf{l}^B$ of $X_\text{syn}[:, B]$. To solve the nearest neighbour problem, we use a simple brute force approach using the Gower coefficient~\cite{gower} to measure the distance between records. We choose this distance since it naturally supports inputs with both categorical and numerical attributes. For categorical attributes, this distance is 1 in the case of a match (or if the two values are both NaNs\footnote{Of the three possible ways in which NaNs can be compared (``NaNs are equal to anything'', ``NaNs are equal to nothing'', and ``NaN is equal to NaN'') considering only ``NaNs equals to NaNs''  gives a broader distribution of distances, which helps identifying close-by records, and gives more effective comparisons in the presence of to suppressed/missing values.}), and 0 otherwise. For numerical attributes the distance is equivalent to the $L_1$ distance, with the values scaled so that $|x_i - x_j| \leq 1$ $\forall x_i, x_j \in \mathbf{x}$. 

The attack procedure is then repeated using the synthetic dataset to establish links between $\Nattacks$ target records drawn from the control set. This results in the two sets of indexes $\mathbf{l}_{\text{control}}^A$ and $\mathbf{l}_{\text{control}}^B$. Finally, a naive attack is implemented to provide a measure probability of finding the correct link by chance. For this $\mathbf{l}_{\text{naive}}^A$ and $\mathbf{l}_{\text{naive}}^B$ are obtained by drawing indexes uniformly at random from the range $[0, n_\text{syn} - 1]$ where $n_\text{syn}$ is the size of the synthetic dataset.

\textbf{Evaluation Phase:}
For each of the $N_A$ targets, we check whether both identified nearest neighbor sets share the same synthetic data record. If they do, the synthetic record allows an attacker to link together previously unconnected pieces of information about a target individual in the original dataset. The attacker scores a success for every correctly established link. The outcome $\mathbf{o}$ of this evaluation is: 
\begin{equation}
    o_i(\mathbf{l}^A_i, \mathbf{l}^B_i) = 
    \begin{cases}
       1 & \text{if} ~ \mathbf{l}^A_i  \cap \mathbf{l}^B_i \neq \emptyset\\
       0 & \text{otherwise}\text{.}\\
    \end{cases}
\end{equation}

This evaluation is performed on the outputs of the three attacks: $(\mathbf{l}^A, \mathbf{l}^B)$ for the attack on training records, $(\mathbf{l}_{\text{control}}^A, \mathbf{l}_{\text{control}}^B)$ for the attack against the control set, and $(\mathbf{l}_{\text{naive}}^A, \mathbf{l}_{\text{naive}}^B)$ for the naive attack. By default, the linkability attack is performed with $k=1$, that is, it considers only the first nearest neighbor. Extending the search to larger values of $k$ helps relax the definition of successful linkage by tolerating a certain degree of ambiguity. This strengthens the attack and is helpful for evaluating synthetic data where no direct one-to-one link between data records might exist.

\subsection{Inference}
For the inference attack, we assume that the attacker knows the values of a set of attributes (the \textit{auxiliary information}) for some target original records. The task of the attacker is to use the synthetic dataset to make correct inferences about some \textit{secret} attributes of the targets. 

\textbf{Attack Phase:}
The core of the inference attack is a nearest neighbor search, similar to what is done in~\cite{Netflix}: for each target record, the attacker looks for the closest synthetic record on the subspace defined by the attributes in the auxiliary information. The nearest neighbour algorithm is the same as that discussed in \Cref{subsec:linkability}. The values for the secret attribute of the closest synthetic record constitutes the guess of the attacker which can then be evaluated for correctness. 
The attack is then repeated against the targets from the control set. Finally, the probability of making a correct inference by chance is measured by implementing a naive inference attack where the attacker's guesses are drawn randomly from the possible values of the secret attribute.

\textbf{Evaluation Phase:}
For evaluation, we say that the attacker has made a successful inference if, for a given secret attribute, the attacker's guess is correct. Comparing the guesses with the true values of the secret in the original data, we count how many times the attacker has made the correct inference. If the secret $s_i$ is a categorical variable, a correct inference requires recovering the exact value. For numerical secrets, the inference is correct if the guess is within a configurable tolerance $\delta$ from the true value:
\begin{equation}
    o_i(s_i, g_i, \delta) =
    \begin{cases}
       \frac{|s_i-g_i|}{s} \leq{\delta} & \text{if $i$ numerical continuous}\\
        1                               & \text{if $s_1 = g_i$ and $i$ categorical}\\
        0                               & \text{if $s_1 \neq g_i$ and $i$ categorical}\text{.}\\
    \end{cases}
\end{equation}
Note that since the same $\delta$ is applied to the main and the control attack, the choice of the particular value of $\delta$ has little impact on the results of the inference analysis.

\section{Experimental Evaluation}\label{sec:experiments}

For practical evaluation, we use three tabular datasets, namely \emph{"Adults"}~\cite{adults}, 1940 US Census~\cite{USCensus} (\emph{"US Census"} in the following), Texas Hospital Discharge Data Public Use Data~\cite{texas} (\emph{"Texas"} in the following). The Adults dataset contains records with 15 mixed-type columns, the Census dataset has 97 mixed-type columns, and the Texas Hospital dataset holds data records with 193 mixed-type columns. The records in the Adults dataset are assigned at random to the training and control dataset with an 80\%-20\% split (39074 training records and 9768 in the control set). The other two datasets are much larger than Adults, and we selected a subset of 50000 records chosen at random for the training dataset, and an equal number of records for the control set. To facilitate synthetic data generation and produce synthetic data with high utility, we consider \emph{subsets} of 37 and 28 columns for the US Census and the Texas dataset, respectively. They are listed in \Cref{app:column_lists}. The choice of which columns are included in this analysis has been based on criteria such as cardinality, number of empty rows, and the presence of other columns which are almost exact duplicates.

\subsection{Linearity of Risk Estimates}\label{subsec:linearity}

This set of experiments demonstrates the ability of our \name to detect and measure various amounts of privacy leaks.

\begin{figure}[t]
\includegraphics[width=0.5\textwidth]{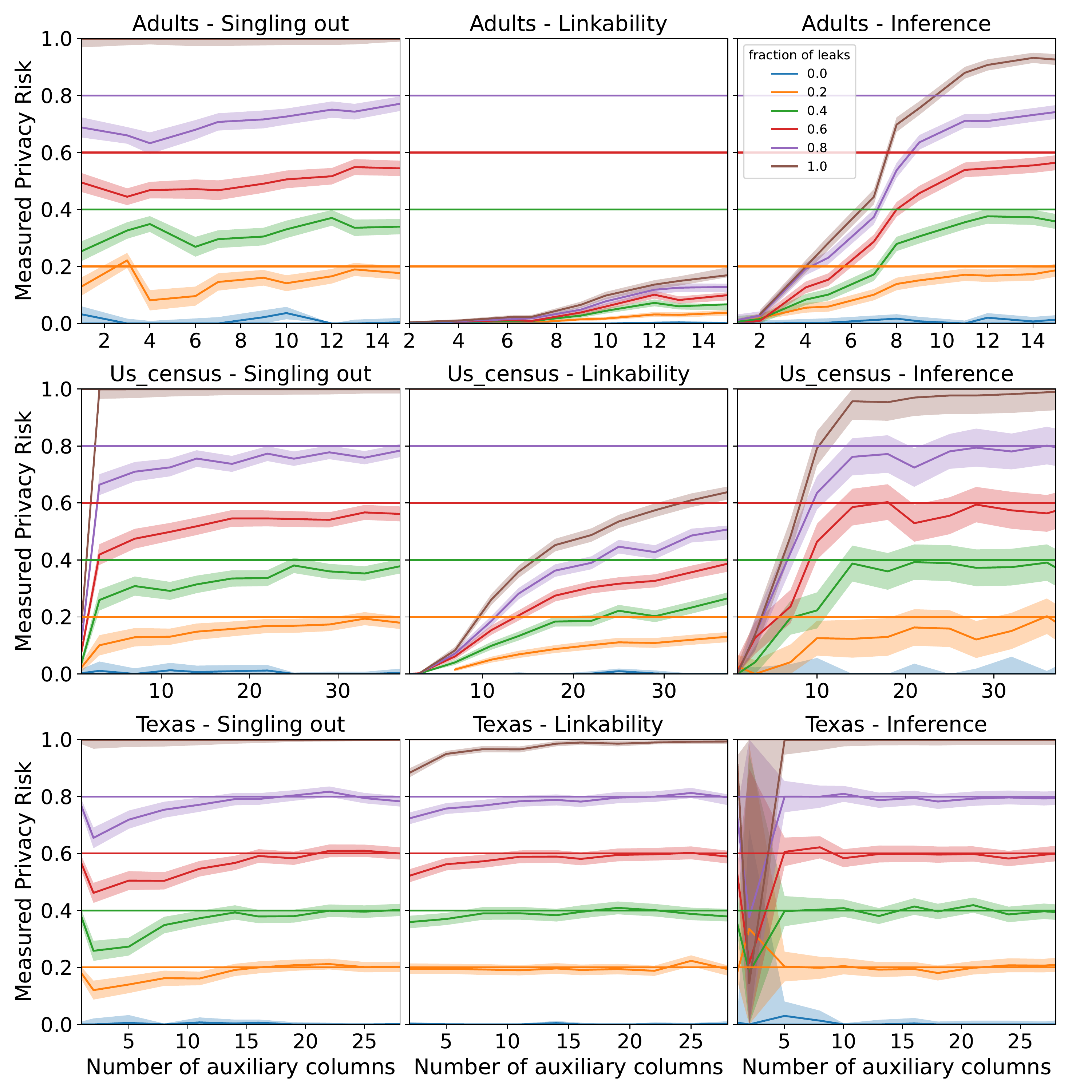}

\caption{Estimated privacy risks as a function of the attacker's auxiliary knowledge for different amounts of artificial privacy leakage. The results from different datasets are arranged vertically: Adults dataset (top), US Census (middle), and Texas (bottom). The horizontal panels show the different privacy risks. From left to right, singling out, linkability, and inference. The shaded areas correspond to the 95\% confidence interval, as measured with the methods presented in \Cref{subsec:architecture}.}
\label{fig:all_risks}
\end{figure}

\setlength{\textfloatsep}{0.1cm}

To assess our privacy risk estimates based on some "ground truth", we first evaluate them without relying on synthetic data since no synthetic data (or anonymization mechanism) can be assumed free of privacy risks. Instead, we split the original dataset in three non-overlapping populations: a training, a control, and a release set of records.
The latter one plays the role of the synthetic dataset in these experiments.
Since the three sets are independent, no information specific to training or control data can be found in the release set.
Based on the training and the release set, we create a ``leaky synthesiser''. 
The leaky synthesizer outputs a dataset $X_\text{syn}^{leaky}$ of size $m$ composed by a fraction $f_l\in[0,1]$ of the training data while the remaining records are taken from the release set. 
The fraction of leaks $f_l$ is the main parameter of this analysis: it represents how many training records ($mf_l$) are directly disclosed.
If $f_l=0$ the leaky synthesizer will exclusively output data records from the release set, thus offering maximum privacy to the training records (noninterference). 
When $f_l=1$ the leaky synthesizer outputs solely training records, so that privacy is maximally violated. 

For these experiments, we set the number of generated predicates $\Nattacks$ to 2000. $\Nattacks$ is one of the free parameters of the analysis and its main influence is to control the statistical uncertainties on the risk estimates. Increasing $\Nattacks$ makes the risk estimate more robust, such that in a practical setting, $\Nattacks$  should be set to yield acceptably small confidence intervals.

The other main parameter of the analysis is the amount of attributes that are used to mount the attacks. This corresponds to the auxiliary information for the linkability and inference attacks, and to the number of attributes in the singling-out predicates in the multivariate singling out case. For brevity, in the following we refer to this parameter simply as auxiliary information for all cases. 
Increasing auxiliary information results in stronger attacks. 
We vary the auxiliary information from the minimum value of one attribute (two in the case of linkability) to all the available attributes (minus one, the secret, for the inference attack), thus modeling increasingly more powerful attacks.

\Cref{fig:all_risks} presents the main results of our experiments. First of all, we observe that by increasing the fraction of artificial privacy leaks---\ie, by replacing a certain amount of data records in the release set by training data records---the privacy risk increases accordingly.
Second, the results highlight that for all three datasets, the measured privacy risks increase linearly with an increased amount of auxiliary attributes known to the attacker. 
As depicted in \Cref{fig:all_risks}, as soon as the auxiliary information covers $10$ to~$20$ attributes, the attack is able to fully exploit the whole fraction of the implanted privacy leaks, and the reported risks saturate to match the fraction of artificially introduced leaks. The linear behaviour of the risk estimates is of great importance when it comes to communicating and interpreting the results of \name:
the framework matches the intuition that the reported privacy leakage grows linearly with the privacy leakage encountered in the data.

When comparing the risk estimates of the different attacks over the three datasets, we observe that the privacy risks in the Texas dataset are higher than in the other two datasets. This is likely due to the many columns with a large number of unique values: 12 (resp., 8) of the 28 columns in the dataset have more than 100 (resp., 1000) unique values. These quasi-identifiers make attacking this dataset particularly easy.
We note that the linearity between the reported risk and the amount of privacy leaks is not fully preserved for the linkability attack in the Adults and the US Census dataset (see the second column, second and third row in \Cref{fig:all_risks}). 
Especially for the Adults dataset, the reported linkability risks is low, $\leq20\%$, even for high fractions of privacy leaks. This result is probably due to the intrinsic properties of the dataset, namely the relatively small number of columns and a prevalence of low cardinality categorical attributes. These factors result in many ambiguous links that the attacker cannot leverage.

Equally important, the reported risks are roughly zero when no privacy leaks occur. 
This "zero-point" of the risk measure derives directly from the fact that the reported risk corresponds to the excess of success by an attacker in comparison to the control baseline---see \Cref{eq:risk}. This property is of great help in practical settings, since it eliminates any possible ambiguity in the interpretation of the measured risks. Any non-zero risk reported by \name has to be seriously inspected.
In particular, since the framework provides a per-column estimate of the privacy risks, it helps identify which attributes, alone or in combinations, exhibit the largest risks (and thereby expose a lot of information on the original data).
As an example, \Cref{fig:inf_per_column} in the Appendix provides an overview on how the inference risk varies depending on which secret attribute the attacker tries to infer.
This information provided by the framework can support the decision-making around synthetic data release: based on the setup, high-risk columns could be removed or perturbed to decrease the privacy risk before releasing the data.

\subsection{Synthetic Data Generation}
\label{subsec:synth_generation}
This section describes how we generate the synthetic data for privacy evaluation using CTGAN~\cite{xu2019modeling}. 
We rely on the original implementation of the algorithm\footnote{\href{https://github.com/sdv-dev/SDV}{https://github.com/sdv-dev/SDV}}, as well as a DP version of the model.\footnote{\href{https://github.com/opendp/smartnoise-sdk/tree/main/synth}{https://github.com/opendp/smartnoise-sdk/tree/main/synth}}

\subsubsection{Synthetization Setup}\label{subsubsec:synth_setup}
For the synthetization with CTGAN, we use the default hyperparameters as specified in the documentation.\footnote{\url{https://sdv.dev/SDV/user_guides/single_table/ctgan.html\#how-to-modify-the-ctgan-hyperparameters}} 
Training DPCTGAN requires more tuning, as we experimentally verified that no usable synthetic data is obtained when running with default parameters. 
Through a hyperparameter search, we identified the best learning rates for the generator and discriminator to be $10^{-5}$ and $10^{-3}$, respectively. 
For each generator update, the discriminator is updated five times, as proposed in~\cite{wgan}. Moreover, as for CTGAN, we use the log frequency for the categorical levels in conditional sampling (\texttt{log\_frequency} is \texttt{True}). Finally, for all datasets, we use a batch size of 64.

\textbf{DP Hyperparameters:} 
Since DPCTGAN cannot be applied to continuous data, the data must be pre-processed before training the model. For details on the pre-processing and its impact on the DP guarantees, see~\Cref{sec:DP_preprocessing}.
We set the noise multiplier in the DPCTGAN training to $0.5$ and the clip norm to $10$. 
With these hyperparameters, we run DPCTGAN for different numbers of epochs (10, 20, 50, 100, 200, respectively) and select the synthetic dataset with the highest utility (see \Cref{subsec:utility} for an overview on how the utility is computed). 
For Adults and US Census the highest utility data results from 100 training epochs, while for Texas 10 epochs yield the best utility.
The rightmost column in \Cref{tab:privacy_utility} and \Cref{fig:utility_scores} in the Appendix provide an overview on the utility of each generated dataset.
The achieved privacy levels depend on the batch size, on the noise multiplier and on the number of training epochs, and are $\varepsilon=32$ for Adults, $\varepsilon=28$ for US Census and $\varepsilon=11$ for the Texas dataset. 
Note that, even though from a theoretical point of view such large values for the privacy budgets might not yield meaningful privacy guarantees, it has been shown that the use of DP, even in the large $\varepsilon$ regime, still results in increased privacy protection in practice~\cite{DPAdversaryBounds}.

\subsubsection{Measuring Utility}\label{subsec:utility}

A utility score between 0 and 100 is attributed to each synthetic dataset as a measure of its data quality. Such score measures three different aspects of the data: marginal distribution similarity, pairwise dependency similarity, and similarity in terms of query counts. Each aspect is quantified separately and the utility score is obtained by averaging them.
The marginal distribution similarity is measured using the Jensen-Shannon divergence~\cite{fuglede2004jensen} between original and synthetic data if the variable is categorical, or the Kolmogorov-Smirnov statistic~\cite{massey1951kolmogorov} for continuous variables. If the data consists of multiple variables, the marginal score is an average over the single variable scores.
The pairwise scores measure how similar correlation, correlation ratio, or mutual information are for a pair of variables in the original and synthetic data. The score is computed based on the absolute error between the original and synthetic statistics. If more than one pair of variables is examined, the pairwise score is averaged over all pairs.
For the query counts similarity, we create a number of random queries and measure how many rows in the original and synthetic data match each query. We then compute the score as the correlation coefficient between the original and synthetic query response sizes. \Cref{fig:utility_scores} in \Cref{app:additional_material} summarizes the utility scores obtained for each dataset and synthetization algorithm. Note how the DP version of the same algorithm gives lower utility.

\subsection{Evaluating Privacy Risks of Synthetic Data}
\label{subsec:synth_eval}
In this set of experiments, we apply \name to the evaluation of the three privacy risks on the generated synthetic data. 

\subsubsection{Evaluation Setup}\label{subsubsec:setup}
For a comprehensive evaluation of the privacy risks of the synthetic data we explore a variety of settings for the attacks algorithms, similar to the experiments of \Cref{subsec:linearity}. In particular, different amount of auxiliary information is evaluated. For each setting, we use $\Nattacks=2000$, repeat the analysis twice and then average the results over both runs. 

For the singling out evaluation we execute both the univariate and multivariate algorithm. For the latter, we repeat the attack with a number of attributes varying between 3 and~12. The final privacy risk is derived from the most successful attack. For linkability different amounts of auxiliary information are considered, ranging from two columns to the maximum number of columns~$d$ in the datasets. For each of the setups, the privacy risks are computed by considering the $k$ nearest neighbors within the synthetic dataset with $k \in \{1, 2,\ldots, 10\}$. Considering $k>1$ neighbours makes it possible to compensate for the fact that synthetic data might not keep one-to-one links between data records, as described in~\Cref{subsec:linkability}. To evaluate inference we mount a set of attacks against each column in the dataset using between 1 and $d - 1$ other columns as auxiliary information.

In total, the above settings result in 120, 480, and 3460 computed privacy estimates for the singling out, linkability, and inference risks, respectively, each corresponding to a different combination of attack settings, dataset, and generative model.

\subsubsection{Quality cuts}
In the following, we present the results only for those settings for which the main privacy attack is stronger than the random baseline (see~\Cref{subsec:architecture}, \emph{valid attacks}). The fraction of such settings is presented in~\Cref{tab:fraction_successful_attacks} for the synthetic datasets generated by CTGAN and DPCTGAN. Furthermore, we exclude attack runs from the results where the success rate $r_{c}$ in the control attack is greater than $0.9$, \ie, attacks where the population-level privacy leakage is very high. In such cases, any additional privacy risk induced through the synthetic dataset that appears when $r_t \geq r_c$ must be relatively small. 
Accurately measuring such small effects often requires larger values of $\Nattacks$ than those used in our analysis due to the underlying statistical randomness. 
Applying these criteria led to excluding approximately $7\%$ and $22\%$  of the results on the inference analysis for the Texas and US census datasets, respectively. Results of the singling out and linkability attacks, and any results for the Adults dataset are not affected by this cut.

\begin{table}[t]
\small
\centering
\begin{tabular}{ccccc}
\toprule                           
\emph{Dataset} & \emph{Method} &  \textbf{Singling out} & \textbf{Linkability} & \textbf{Inference}      \\
\midrule
\multirow{2}{*}{\emph{Adults}}    & \emph{CTGAN} &    100\%       &  68\%    & 86\%     \\
                                  & \emph{DPCTGAN} &   90\%      &  52\%       & 77\% \\
\hline
\multirow{2}{*}{\emph{Texas}}     &\emph{CTGAN} &    84\%      &  79\%       &86\% \\
                                  & \emph{DPCTGAN} &  24\%     &  41\%       &60\%  \\
\hline
\multirow{2}{*}{\emph{US Census}} & \emph{CTGAN} &    90\%      &  82\%       &88\% \\
                                  & \emph{DPCTGAN} &   88\%      &  73\%       &85\% \\
\bottomrule
\end{tabular}
\caption{Fraction of risk estimates which resulted in a valid attack, \ie, for which the success rate of the main privacy attack is above that of the random baseline algorithm. Percentages are given over the 120, 480, and 3460 computed privacy estimates for singling out, likability, and inference, respectively.
\label{tab:fraction_successful_attacks}}
\end{table}
\setlength{\textfloatsep}{-0.08cm}

\begin{figure*}
\centering
\includegraphics[width=0.7\textwidth]{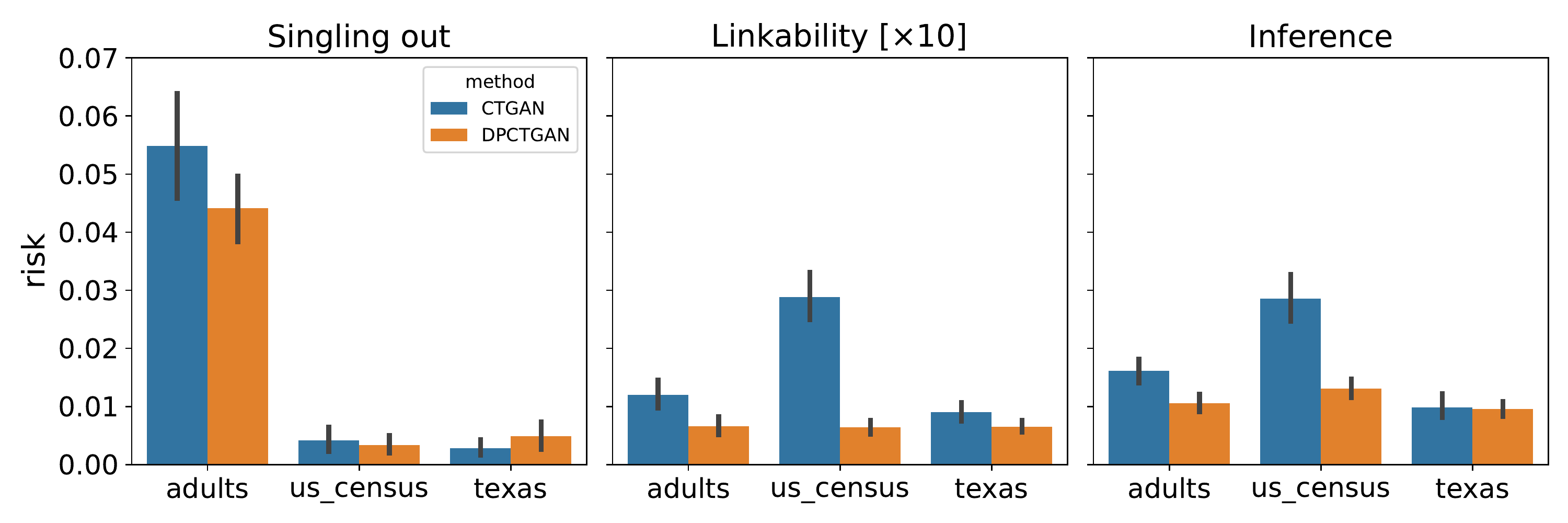}
\caption{Overview of the privacy risks scores for different datasets and synthetization methods. For the sake of visualization, the linkability risk score has been increased by a factor ten. \label{fig:risk_reduction}}
\end{figure*}

\begin{table*}[t]
\centering
\begin{tabular}{cccccc}
\toprule
\emph{Dataset} & \emph{Method} &  \textbf{Singling out} & \textbf{Linkability} & \textbf{Inference}  &           \textbf{Utility} \\
\midrule
\multirow{2}{*}{\emph{Adults}} & \emph{CTGAN} &  0.0548 $\pm$ 0.0134 &  0.0012 $\pm$ 0.0003 &  0.0161 $\pm$ 0.0031 &  88 \\
          & \emph{DPCTGAN} &  0.0441 $\pm$ 0.0084 &  0.0007 $\pm$ 0.0002 &  0.0106 $\pm$ 0.0021 &  79 \\
\hline
\multirow{2}{*}{\emph{Texas}}     & \emph{CTGAN} &  0.0028 $\pm$ 0.0020 &  0.0009 $\pm$ 0.0002 &  0.0098 $\pm$ 0.0030 &  88 \\
          & \emph{DPCTGAN} &  0.0049 $\pm$ 0.0037 &  0.0007 $\pm$ 0.0002 &  0.0095 $\pm$ 0.0019 &  71 \\
\hline
\multirow{2}{*}{\emph{US Census}} & \emph{CTGAN} &  0.0042 $\pm$ 0.0032 &  0.0029 $\pm$ 0.0006 &  0.0286 $\pm$ 0.0057 &  74 \\
          & \emph{DPCTGAN} &  0.0033 $\pm$ 0.0022 &  0.0006 $\pm$ 0.0002 &  0.0131 $\pm$ 0.0023 &  63 \\
\end{tabular}
\caption{Measured risks and utility scores for the different attack types, datasets, and generative models. The reported risks are obtained by averaging over all the attack settings (see \Cref{subsubsec:setup}) and the reported uncertainties shows the 95\% confidence interval computed via bootstrapping.\label{tab:privacy_utility}}
\end{table*}

\subsubsection{Evaluation Results}

We evaluate the privacy risks for all three attacks on the synthetic datasets generated by CTGAN and DPCTGAN. \Cref{fig:risk_reduction} shows that, for the Adults and the US Census datasets, the synthetic data generated with DP guarantees exhibit smaller privacy risks than their non-DP counterpart. When considering the ratios of different privacy risk scores between the non-DP and DP synthetic data for Adults and US Census, we observe that the ratio for inference (Adults: 1.52, US Census 2.18) and linkability (Adults: 1.82, US Census: 4.53) are higher than for singling out (Adults: 1.24, US Census 1.25). This suggests that DP mitigates the former two privacy risks to a higher extent than the singling out risk.
We attribute the cause of such difference to the fact that, regardless of how the generative model is trained, the values of the categorical attributes in the synthetic dataset are sampled from those present in the original dataset (see \cite{golden_ferrari} and the discussion in \Cref{sec:DP_preprocessing} for details). Of the three privacy attacks, singling out is the one which puts the greater emphasis on the exact values of these attributes.

However, for the Texas dataset, despite having a smaller DP level $\varepsilon$, the mitigation of the privacy risk offered by DP is not as effective. We suspect that this is either a result of the DP guarantees not being protective enough, or of the pre-processing described in~\Cref{sec:DP_preprocessing} which preserves too detailed information on the respective categorical diagnostic codes in the dataset, or of a combination of both effects together with the intrinsic properties of that particular dataset (number of attributes, their distributions and their ranges).

Furthermore, we see that risks reported for linkability are low over all three datasets. This supports the common intuition that synthetic data indeed does not preserve one-to-one links between data records. 

We also display the privacy risks as a function of the auxiliary information available to the attacker.
For linkability and inference risk, the amount of auxiliary information is given by the number of auxiliary columns, whereas for singling out, the amount of auxiliary information corresponds to the number of attributes used to create the predicates.
\Cref{fig:risks_vs_aux} in \Cref{app:additional_material} shows the respective risks for each attack.  

\Cref{fig:linkability_num_neighbors} in \Cref{app:additional_material} depicts the linkability risk not only as a function of the number of auxiliary columns, but also as a function of the number $k$ of nearest neighbors used to determine the linkage for both CTGAN and DPCTGAN. Although a clear trend can be observed, showing an increase in the reported linkability risk as the number of auxiliary information columns grows, the effect of including a higher number of neighbors in the nearest neighbor search is less consistent. This further supports the thesis that synthetic data indeed helps thwart linkability attacks.

As seen in \Cref{subsec:utility}, synthetic data generated with CTGAN has more utility than those from DPCTGAN. 
\Cref{tab:privacy_utility} compares the changes in the measured privacy risks among data generated with the respective generator models. In both the Adults and US Census datasets, we observe that the data generated with DP guarantees exhibit lower privacy risks.
This might also be explained by synthetic data with higher utility more closely matching the original data, thereby exposing an increased risk to the various privacy attacks. In the Texas dataset, however, the privacy risks are similar for both CTGAN and DPCTGAN and they remain low even if CTGAN offers better utility.

\subsection{Quantitative Comparison with Previous Work}\label{sec:comparison}

In this experiment, we compare our methods with~\cite{groundhog} which we refer to as \textit{Synthetic Data Release} (SDR~\cite{SDR_code}). 

To evaluate privacy risks, we rely on the "leaky synthesizer" methodology presented in \Cref{subsec:linearity}.
We instantiate six different leaky synthesizers, each one corresponding to a different $f_l$ between $0$ and $1$ and use SDR to evaluate their linkability and inference privacy risks, leaving all the analysis settings unchanged from those presented in~\cite{groundhog}. The analysis uses the Texas dataset included in the SDR code release~\cite{SDR_code}, which contains a smaller set of columns than those used in the rest of this paper (18 instead of 28). The leaky datasets created for evaluating SDR are saved to disk, and are then evaluated using \name. For this evaluation, we vary the number of auxiliary information used to mount the attack between 10 and the maximum number of available attributes. For the linkability attack, we consider linkability at the level of the closest neighbour only (see \Cref{subsec:architecture}), as done for the experiments presented in \Cref{subsec:linearity}. 

\begin{figure}[t]
\includegraphics[width=0.5\textwidth]{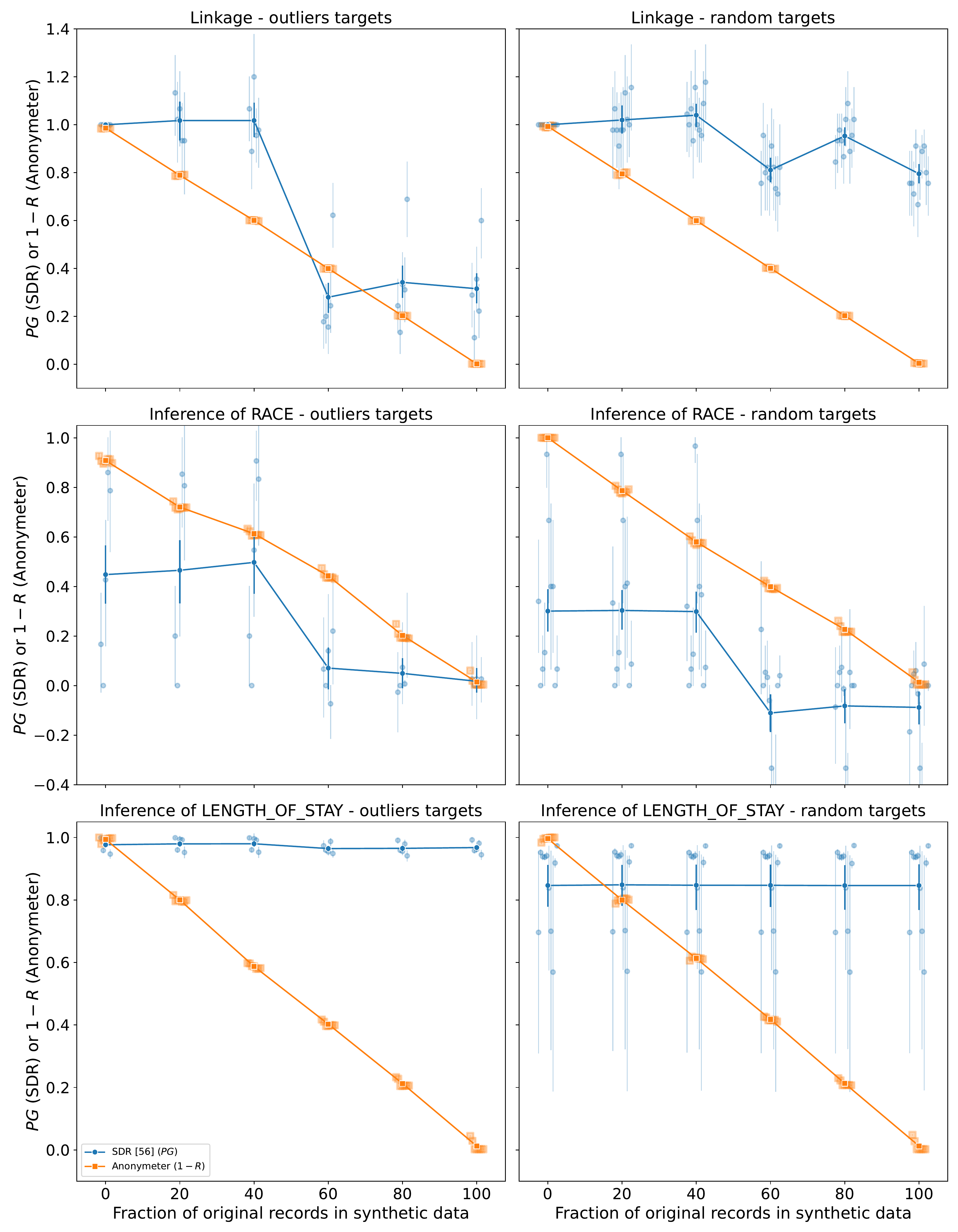}
\caption{Privacy gain $PG$ (blue) and $1-R$ (orange) as a function of the amount of privacy leaks. The plots on the left side are obtained from the analysis of hand-selected target records (the same as those chosen in~\cite{groundhog}), those on the right uses 10 randomly chosen targets. The blue points report the $PG$ of each target. The orange squares present the privacy risk estimated for various amounts of auxiliary information. Although the two metrics are reported on the same axis, care must be taken when comparing their values. See text for details.}
\label{fig:compare_privacy_gains}
\end{figure}
\setlength{\textfloatsep}{0.1cm}

As described in \Cref{sec:background}, the metric used in SDR  is the Privacy Gain (PG). In our work, we instead use the privacy risk (R). These two quantities are defined differently and therefore care must be taken when comparing them: in particular, one should not assume that $PG = x$ has the same meaning as $1-R = x$. However, both serve the same purpose and one should expect that a decreasing $1-R$ would be matched by a reduced $PG$—and vice versa, so that a high level comparison is indeed possible. \Cref{fig:compare_privacy_gains} presents the results of such a comparison between SDR and \name for different values of $f_l$.

For the linkability attack (top row), the success of SDR is strongly influenced by the selection of the target records. For randomly selected targets, the $PG$ is always high and shows only a mild dependency on $f_l$. When the evaluation targets selected outlier points (the same as in \cite{groundhog}) we see a step-like variation of $PG$ with the $f_l$: $PG \sim 1$ for $f_l \lesssim 0.5$ and $PG \sim 0$ for $f_l \gtrsim 0.5$.
For the inference attack, the performance of SDR varies greatly depending on whether the secret that is being guessed is categorical (\texttt{RACE}, middle row) or continuous (\texttt{LENGTH\_OF\_STAY}, bottom row). In the former case, $PG$ goes from $\sim 0.4$ to $\lesssim 0$ across the range of privacy leaks, again showing a sharp drop at $f_l\sim 0.5$. When the secret is continuous however, PG remains $\sim 1$ for all values of $f_l$, even if the synthesizer leaks the entire training data.

In contrast, the results of \name, for both inference and linkability, vary linearly with $f_l$ between $1-R=1$ for $f_l=0$ and $1-R=0$ for $f_l=1$, as expected. This result aligns with the experiments presented in \Cref{subsec:linearity} and confirms that \name outputs ``well behaved'' privacy metrics.

The results support our claim that \name is a robust way to capture various degrees of privacy leaks. Perhaps more importantly, \name never fails to report $R>0$ when privacy leaks are present. \name also offers better scalability to large datasets than SDR. This is because SDR requires training dozens of models and generating thousands of synthetic datasets, which restricts the practical usability of the method to datasets with a maximum of tens of thousands of data records. For example, for each of the 6 leaky synthesizers, SDR created 2550 and 1250 synthetic datasets to analyze inference and linkability, respectively. In contrast, Anonymeter only requires one realization of the synthetic dataset and can evaluate the privacy of large synthetic datasets with millions of rows within less than one day of computing time using rather inexpensive general-purpose virtual machines with 64 virtual CPUs. For the analysis just described, SDR was 10 times slower than \name in wall clock time. Such performance difference would have been even more pronounced for a ``real'' synthesizer using time-consuming training algorithms.
\section{Discussion and Future Direction}

The evaluation of \name on singling out, linkability, and inference risks highlights the effectiveness of our framework to provide a coherent assessment of legally meaningful privacy metrics. Not only does \name allow us to analyze general privacy leakage as a function of the attacker's power, but at the same time it helps identify concrete privacy violations in the synthetic datasets. 
In particular, \name significantly outperforms existing frameworks for privacy evaluation of synthetic data in both the detection of privacy leakage and computational complexity.
This is a crucial step on the way towards leveraging the full potential of using synthetic data while keeping track of the privacy implications.

Moreover, the modular nature of \name facilitates the future integration of new and potentially stronger attacks for evaluating the three privacy risks analyzed in this work. Privacy attacks that evaluate other aspects of privacy, such as membership inference, can also be integrated. This flexibility allows the framework to adapt to and to meet future requirements from emerging and changing privacy regulations.

Another advantage of \name is that it separates the evaluation of the success rate of the privacy attacks from the calculation of the reported privacy risks. Due to the statistical nature of \name's risk quantification phase (where each attack simply yields a boolean array), the privacy risk is deduced from the main attack and the baselines, which provide the necessary context for turning attack success into expressive privacy risks.

Since \name treats the synthetic data generation mechanism as a black box and solely utilizes the generated dataset, the framework can be used on datasets anonymized with other methods. \name can even be applied to an original dataset to identify individual data records with high privacy risks.
This assessment can, among others, serve as a pre-filtering mechanism to identify---and, for example, remove---high-risk data records before training a generative model on the original data. This can lead to reduced privacy risks for the generated synthetic dataset.

In this work, we apply \name solely to perform privacy assessments of entire datasets. 
However, the framework can be directly applied to quantify the privacy risks associated with a particular individual (or subgroup of individuals) in the dataset.
Therefore, the generation of guesses in the attack phase just has to specify the respective individual(s) instead of using random targets.
To provide a more fine-grained risk assessment over the entire dataset, the target selection in our framework could also rely on identifying targets with high privacy risks and selecting these for generating the guesses.
This would help approximate the upper bound on privacy leakage in the dataset more closely than an assessment over randomly chosen targets.

Future work could also look into the extension of \name for assessing additional trustworthiness aspects in synthetic datasets, such as their fairness, or existing biases towards certain subgroups of individuals,
thereby supporting a joint assessment of these different aspects together with utility and privacy of the data.

\section{Conclusion}

Synthetic data has the potential to mitigate existing tensions between the need to share and utilize sensitive datasets and the privacy concerns of the individuals whose data is included in these datasets.
The fact that the actual privacy leakage in such datasets is hard to quantify, hinders leveraging the high potential of the data.
To close this gap, we propose \name, a statistical framework for jointly quantifying different privacy risks in synthetic datasets.  
Within this framework, we implement concrete attacks to measure the privacy risks of singling out, linkability, and inference---the three risks that anonymization methods must mitigate to be legally compliant with existing privacy legislation. Our work is the first to propose practical attacks directly measuring the singling out and linkability risks posed by the release of a synthetic dataset. The experiments we conduct highlight that \name is able to report privacy risks in a coherent and fine-grained manner, making the framework a valuable resource for identifying privacy leakage and quantifying the corresponding risks. We also demonstrate that \name significantly outperforms prior work both in finding privacy leaks as well as in usability.
By making \name available as an easy to use open source library, we hope to contribute to a more mature and privacy-aware usage of synthetic data. In particular, we want to help avoid dangerous situations in which, due to the lack of usable privacy analysis, synthetic data is simply ``trusted'' to be private.

\begin{acks}
We thank the authors of~\cite{groundhog} for making their work easy to reproduce and to interact with. We also thank the Synthetic Data Vault and Smartnoise teams for providing the synthetic data software used in this work. We thank Nicola Vitacolonna for the great help and support in going through the revision process.
\end{acks}

\bibliographystyle{ACM-Reference-Format}
\bibliography{bibliography}

\appendix
\section{Utility of synthetic data and size of control dataset }\label{app:control_ablation}

\name requires the user to set aside a small portion of the original data for the control dataset. We perform an ablation study to determine how the reduced size of the training set impacts utility. In this experiment, we synthesize the three datasets using CTGAN and the same setup as described in \Cref{subsubsec:synth_setup}. For each of the three datasets, the training set is made of a varying fraction (from $10\%$ to $99\%$) of original records. A synthetic dataset with the same number of samples as the original data is then created starting from the different sized training sets, and its utility is evaluated using the method proposed in \Cref{subsec:utility}, comparing it to the entire original dataset. The entire process is repeated three times and the results aggregated to account for the randomness involved in both training and sampling a generative model.

\begin{figure}[htbp]
\centering
\includegraphics[width=0.5\textwidth]{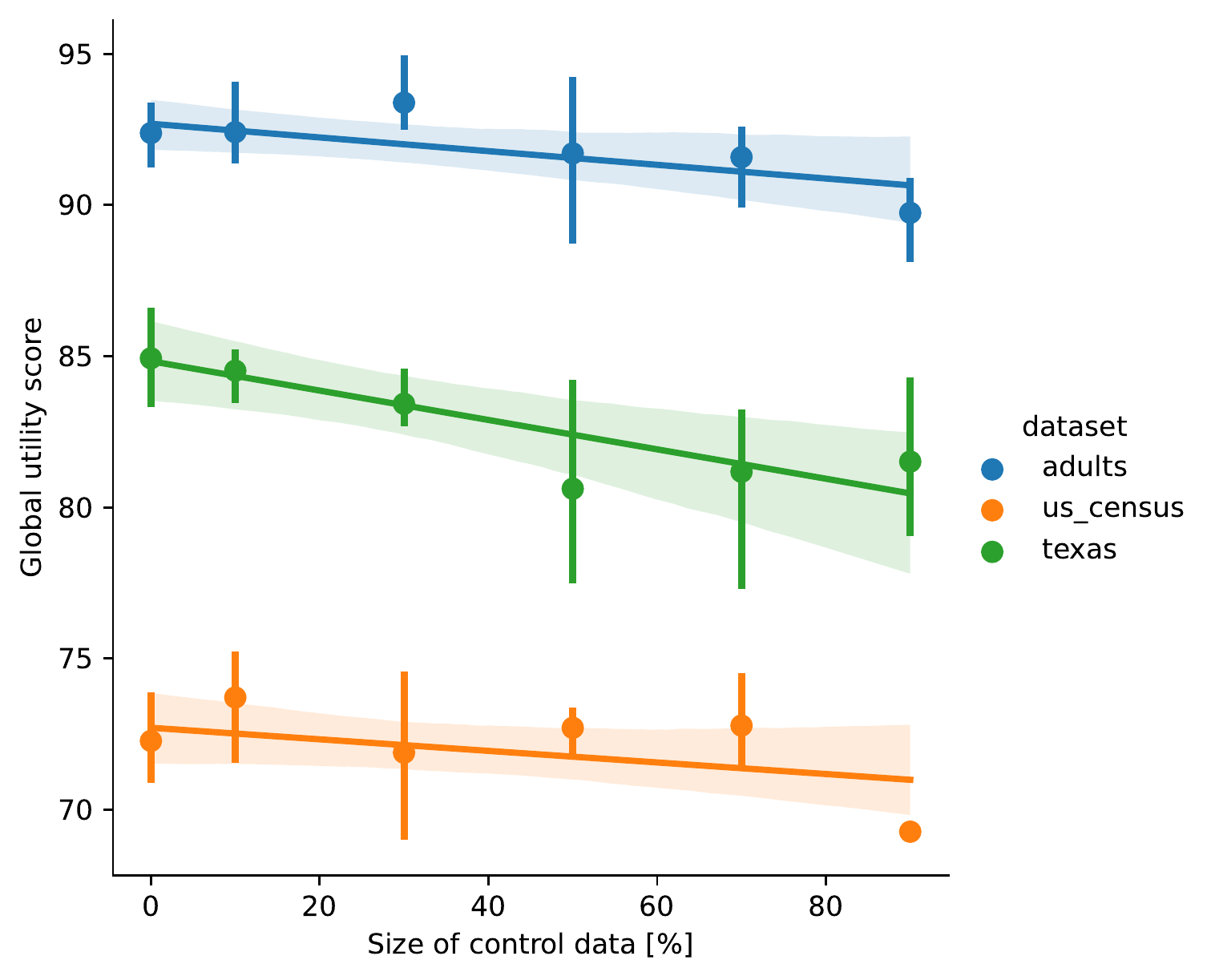}
\caption{Measured utility as a function of the size of the control dataset for the Adults (blue), US Census (orange), and Texas (green) dataset.\label{fig:control_size}}
\end{figure}

The results are presented in Figure~\ref{fig:control_size}. As expected, there is a degradation in utility with decreasing size of the training data. However, this effect is mild, especially if one considers that the control dataset needs to contain just as many records as the number of attacks ($N_{attacks}$) that will be performed during the evaluation. A few thousands records in the control set is generally enough to have small statistical uncertainties on the measured risks. The $N_{attacks} = 2000$ as used in this work corresponds to a relative size of the control set of $4\%$, that is, $96\%$ records can be used for training. As visible in Figure~\ref{fig:control_size}, there is no measurable difference in utility for such small control size fractions.

To conclude, our ablation study shows that, if the control dataset is relatively small (\ie, less than 20\% of the total), utility is hardly affected. The requirements of \name to have a control dataset does not represent a limitation to its practical use. The situation might be different in case of very small datasets or if the focus is on rare records (for example, in the case of anomaly detection). For both these cases, however, the use of synthetic data can be very problematic.

\section{Modeling singling out success for different population sizes}\label{app:so_correction}

The success of a singling out attempt depends on the size of the target population. A birth date will likely single out an individual in a sample of 365 people, but not when considering all the inhabitants of a big city. In this section we present a model that allows us to compare the singling out success rates $r_\text{train}$ and $r_\text{control}$ in the general case in which the training set is larger than the control set.

More formally, let $N_{\text{train}}$ and $N_\text{control}$ be the number of records in the training and control set, respectively (generally $N_\text{control} < N_\text{train}$). The attacker builds $\Nattacks$ predicates generated from the synthetic dataset: $m_\text{train}$ of these isolate a record in the train dataset, and $m_\text{control}$ isolate a record in the control dataset. We want to derive $\tilde{m}_\text{control}$, the number of predicates that \textit{would} isolate records in the control set, if the control set had $N_\text{train}$ many records.

The probability that a predicate isolates a record in a population depends on the weight $w$ of the predicate, \ie, the probability that it is satisfied by exactly one record sampled from the data generating distribution $\mathcal{D}$. In particular, for a population of size $n$, this probability is given by~\cite{PredicateSinglingOut}:

\begin{equation}\label{eq:so_prob}
P(w, n) = nw(1-w)^{n-1} \text{.}  
\end{equation}

As $w$ depends on the distribution $\mathcal{D}$, which is unknown, Equation~\ref{eq:so_prob} cannot be used directly to compensate for the different size of the control set. However, the predicates we evaluate are fixed (generated by the attack) and so are their weights, since they are fully defined by the query and statistic of the population of interest. Moreover, these predicates single out in the synthetic dataset (which has size $N_\text{train}$), so their weights should be small, and of the order of $1/N_\text{syn}$. To focus on this low-weight regime we integrate $P(n, w)$ in the range $[0, w_\text{eff}]$. Our model then becomes:
    
\begin{equation}\label{eq:so_model}
S(n) = A\int_{0}^{w_\text{eff}} P(w, n) dw  \text{.} 
\end{equation}

The parameter $w_{\text{eff}}$ can be interpreted as the "effective weight" of the set of attack predicates, while $A$ is a normalization constant to convert the probability into counts. To determine these parameters for the specific dataset and attack predicates, \name evaluates the attack predicates on a set of samples of control records of different sizes, in order to measure how the number of successful singling out predicates changes with the dataset size $n$. The evaluator automatically samples 10 different values for $n$ from range [1000, $N_\text{control}$], and for each it repeats the measurement on five different samples of control records, selected at random without replacement from the control set. This gives enough data points (five for each of the ten values of $n$) to fit our simple two parameter model with confidence. Once the parameters of $S(n)$ have been fitted, we can derive $\tilde{m}_\text{control}$ as:

\begin{equation}\label{eq:so_correction}
    \tilde{m}_\text{control} = \frac{S(N_{\text{train}})}{S(N_\text{control})} m_\text{control}\text{.}
\end{equation}

\begin{figure}[t]
  \includegraphics[width=0.5\textwidth]{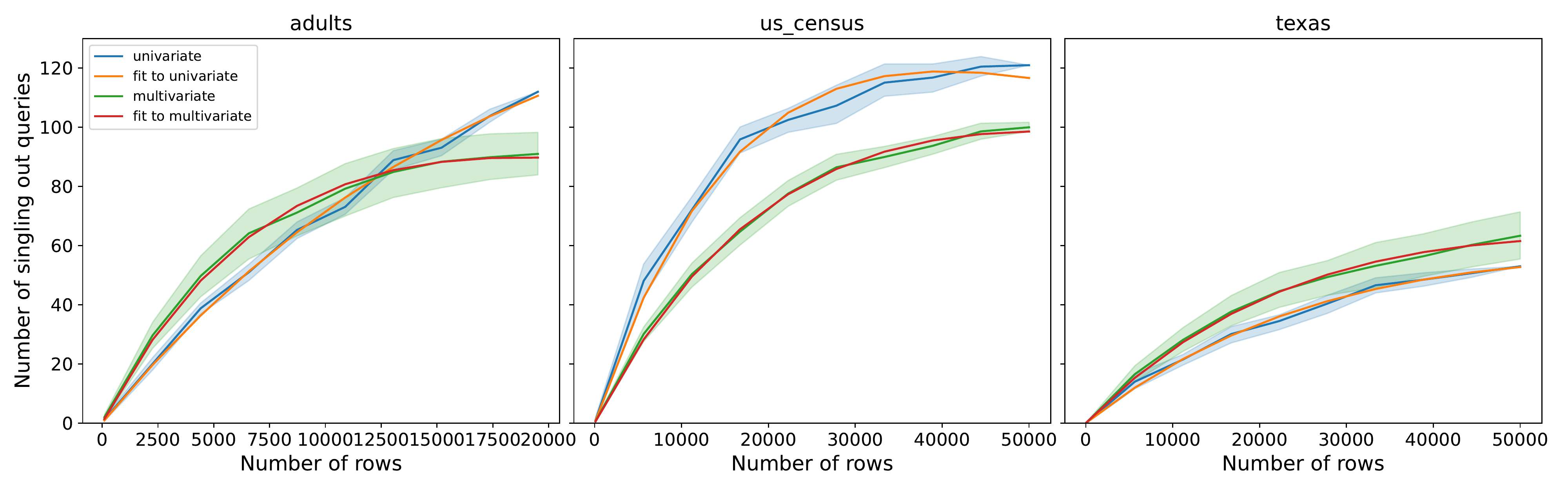}
\caption{Number of successful singling out predicates as a function of the dataset size together with best fit model $S(n)$. The three panels refers to the different datasets: Adults (left),  US census (middle), Texas (right). \label{fig:so_correction}}
\end{figure}

In Figure~\ref{fig:so_correction} we present the results of a set of experiments done to evaluate the quality of our statistical model of the singling out success (Equation~\ref{eq:so_model}) which underpins our correction factor (Equation~\ref{eq:so_correction}). 
The datasets used for the evaluation are formally introduced in Section~\ref{sec:experiments}. 
In all cases, the original dataset is split into two disjoint sets, called A and B. From set A, $\Nattacks$ singling out predicates are derived using the methods presented above. The attack predicates are then evaluated on different subsets of B. By varying the sizes of these subsets, we can measure how the the number of successful singling out predicates varies with the size of the target population. Figure~\ref{fig:so_correction} shows that the fitted statistical model $S(n)$ (Equation~\ref{eq:so_model}) provides very good agreement with the measured data points. 

\section{Column subsets included in the analysis}\label{app:column_lists}

For the Texas and US Census datasets, the following columns have been used throughout all experiments presented in this work:

\paragraph{Texas dataset:} \texttt{ADMIT\_WEEKDAY}, \texttt{APR\_DRG}, \texttt{APR\_MDC}, \texttt{CERT\_STATUS}, \texttt{COUNTY}, \texttt{ETHNICITY}, \texttt{FIRST\_PAYMENT\_SRC}, \texttt{ILLNESS\_SEVERITY},\\ \texttt{LENGTH\_OF\_STAY}, \texttt{PAT\_AGE}, \texttt{PAT\_COUNTRY}, \texttt{PAT\_STATE}, \texttt{PAT\_STATUS}, \texttt{PAT\_ZIP}, \texttt{PRINC\_DIAG\_CODE}, \texttt{PROVIDER\_NAME}, \texttt{PUBLIC\_HEALTH\_REGION}, \texttt{RACE}, \texttt{RISK\_MORTALITY}, \texttt{SEX\_CODE}, \texttt{SOURCE\_OF\_ADMISSION}, \\
\texttt{TOTAL\_CHARGES}, \texttt{TOTAL\_CHARGES\_ACCOMM}, \texttt{TOTAL\_CHARGES\_ANCIL}, \\
\texttt{TOTAL\_NON\_COV\_CHARGES}, \texttt{TOTAL\_NON\_COV\_CHARGES\_ACCOMM}, \\ \texttt{TOTAL\_NON\_COV\_CHARGES\_ANCIL}, \texttt{TYPE\_OF\_ADMISSION}

\paragraph{US Census dataset:} \texttt{AGE}, \texttt{AGEMARR}, \texttt{AGEMONTH}, \texttt{BPL}, \texttt{CHBORN}, \texttt{CITIZEN}, \texttt{EDUC}, \texttt{ELDCH}, \texttt{EMPSTAT}, \texttt{FAMSIZE}, \texttt{HIGRADE}, \texttt{HISPAN}, \texttt{HRSWORK1}, \texttt{INCWAGE}, \texttt{IND1950}, \texttt{MARST}, \texttt{MBPL}, \texttt{MOMRULE\_HIST}, \texttt{MTONGUE}, \texttt{NATIVITY}, \texttt{NCHILD}, \texttt{NSIBS}, \texttt{OCC1950}, \texttt{POPLOC}, \texttt{POPRULE\_HIST}, \texttt{RACE}, \texttt{RELATE}, \texttt{RELATED}, \texttt{SEX}, \texttt{SPLOC}, \texttt{SPRULE\_HIST}, \texttt{STEPMOM}, \texttt{STEPPOP}, \texttt{SUBFAM}, \texttt{UOCC}, \texttt{WKSWORK1}, \texttt{YNGCH}

\section{Differential Privacy and data pre-processing}
\label{sec:DP_preprocessing}
Since DPCTGAN cannot be applied to continuous data, we encode the continuous columns of all datasets by binning them into 50 equally spaced bins, respectively.
After synthetization, continuous values are obtained from each synthetic bin label by drawing random numbers from a uniform distribution spanning the interval defined by the corresponding bin edges.
Note that even though the data synthetization with DP-CTGAN incorporates DP guarantees, the pre-processing itself does not.
This is due to the ranges from which the bins are derived being obtained from the private data itself. 
This problem is common to many open source data synthesizers~\cite{groundhog}. 

We note, however, that the problem is not limited to the continuous domain and affects also categories. Categorical data is also sampled from distributions whose support is derived from the input dataset itself. This leads to a violation of the DP guarantee which follows from the bad bins theorem~\cite{golden_ferrari}: \textit{``There does not exist a differentially private mechanism that only returns elements from the active domain.''}. This means that one cannot obtain DP for synthetic data if the synthesizer can only generate values that are actually present in the original data instead of also allowing for the presence of values which are possible but not present in the original dataset. Fixing all these issues is an interesting research topic, but out of the scope of this paper. The reader should however keep in mind that DP is not complete (and hence devoid of theoretical meaning).

\section{Additional Experimental Results}\label{app:additional_material}

This section collects additional experimental results supporting the discussions presented in Sections \ref{subsec:singling_out}, \ref{subsec:linearity}, \ref{subsec:synth_eval}, and \ref{sec:comparison}. 

\begin{figure}[tbh]
\centering
\includegraphics[width=0.35\textwidth]{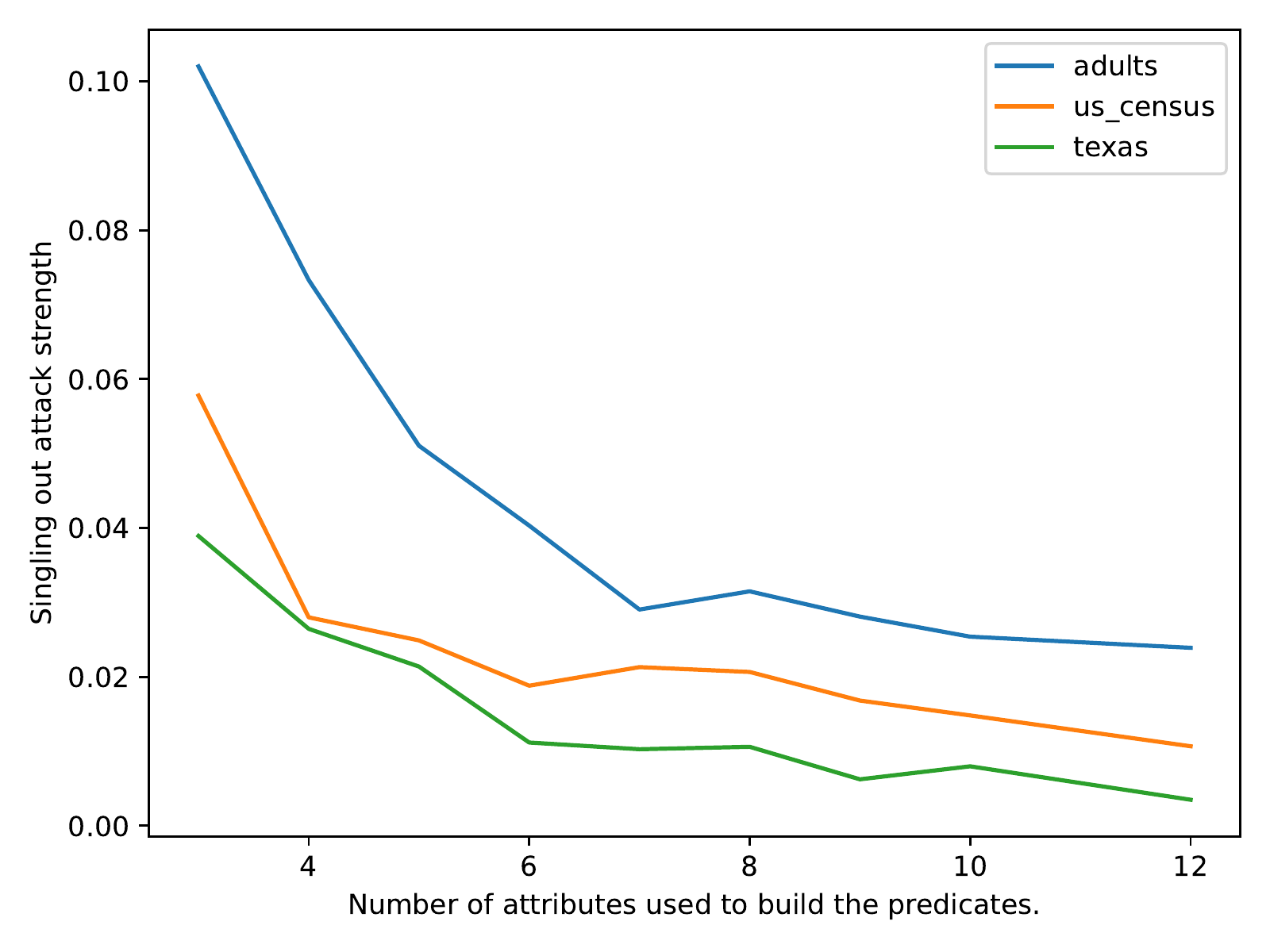}
\caption{Strength of the multivariate singling out attack as a function of the number of attributes used to create the predicates. The singling out attack is stronger when it uses fewer attributes. As the number of attributes used to create the predicates increases, the likelihood of singling out by chance also improves, making the baseline attack more and more effective. This results in a decreasing strength of the multivariate singling out attack for a higher number of attributes.}
\label{fig:so_strength}
\end{figure}

\begin{figure}[tbh]
\centering
\includegraphics[width=0.35\textwidth]{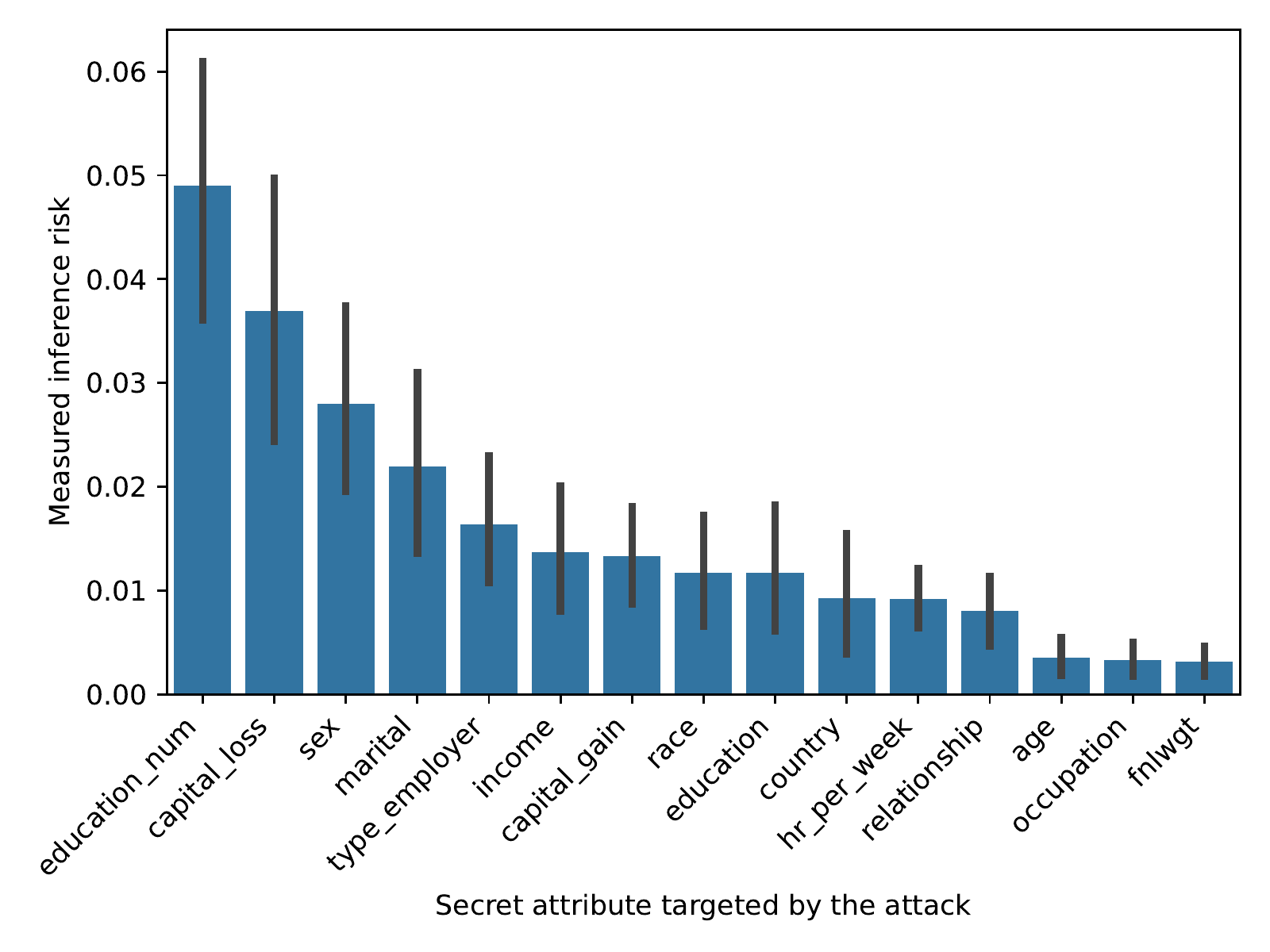}
\caption{Inference risk for different attributes in the Adults dataset. The synthetic data has been generated with CTGAN from \Cref{subsec:synth_generation}, and each column is attacked by using all the other columns as auxiliary information.}
\label{fig:inf_per_column}
\end{figure}

\begin{figure}[tbh]
\centering
\includegraphics[width=0.35\textwidth]{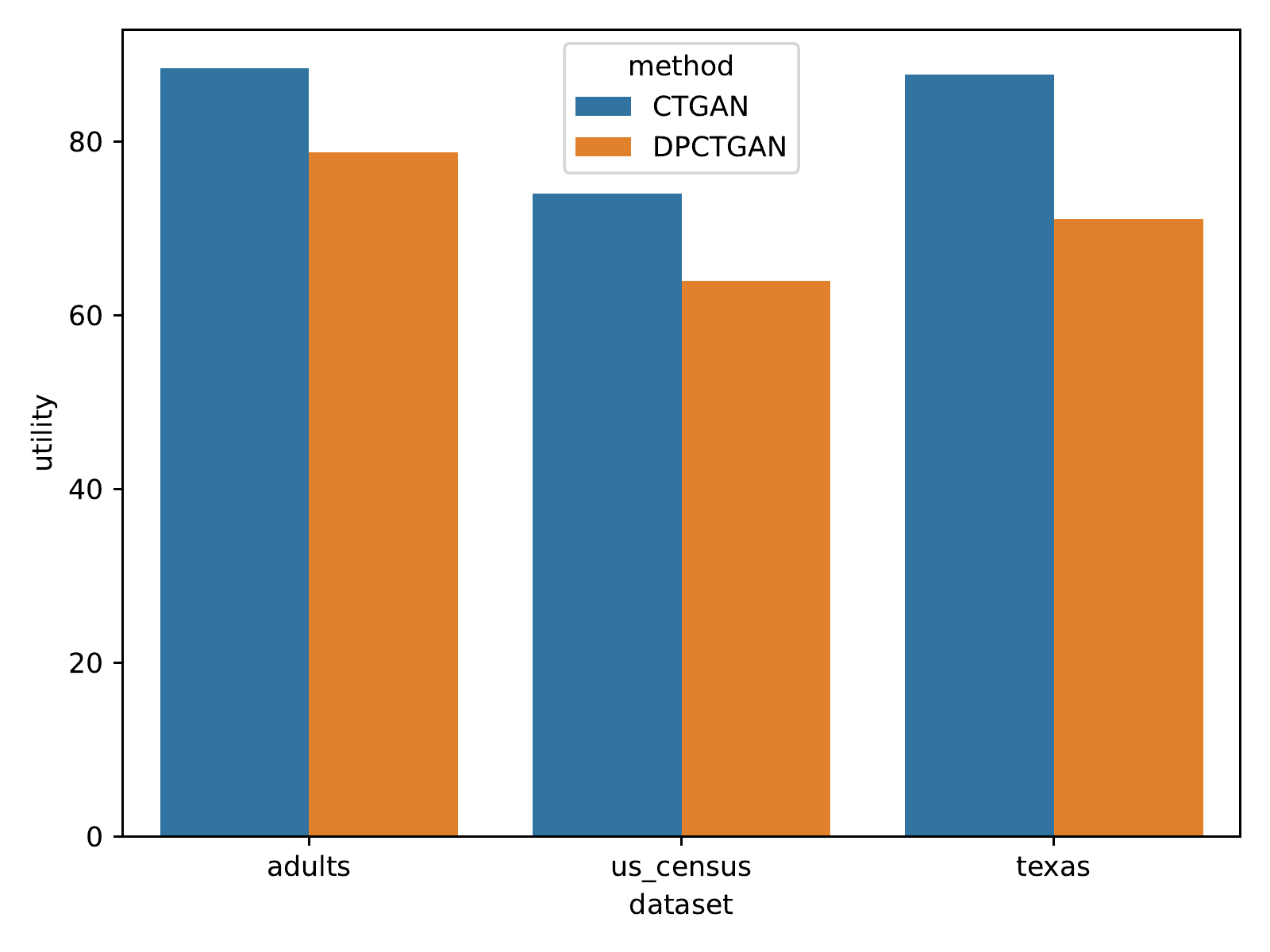}
\caption{Utility scores for the different datasets and synthetization algorithms. \label{fig:utility_scores}}
\end{figure}

\begin{figure}[tbh]
\centering
\includegraphics[width=0.25\textwidth]{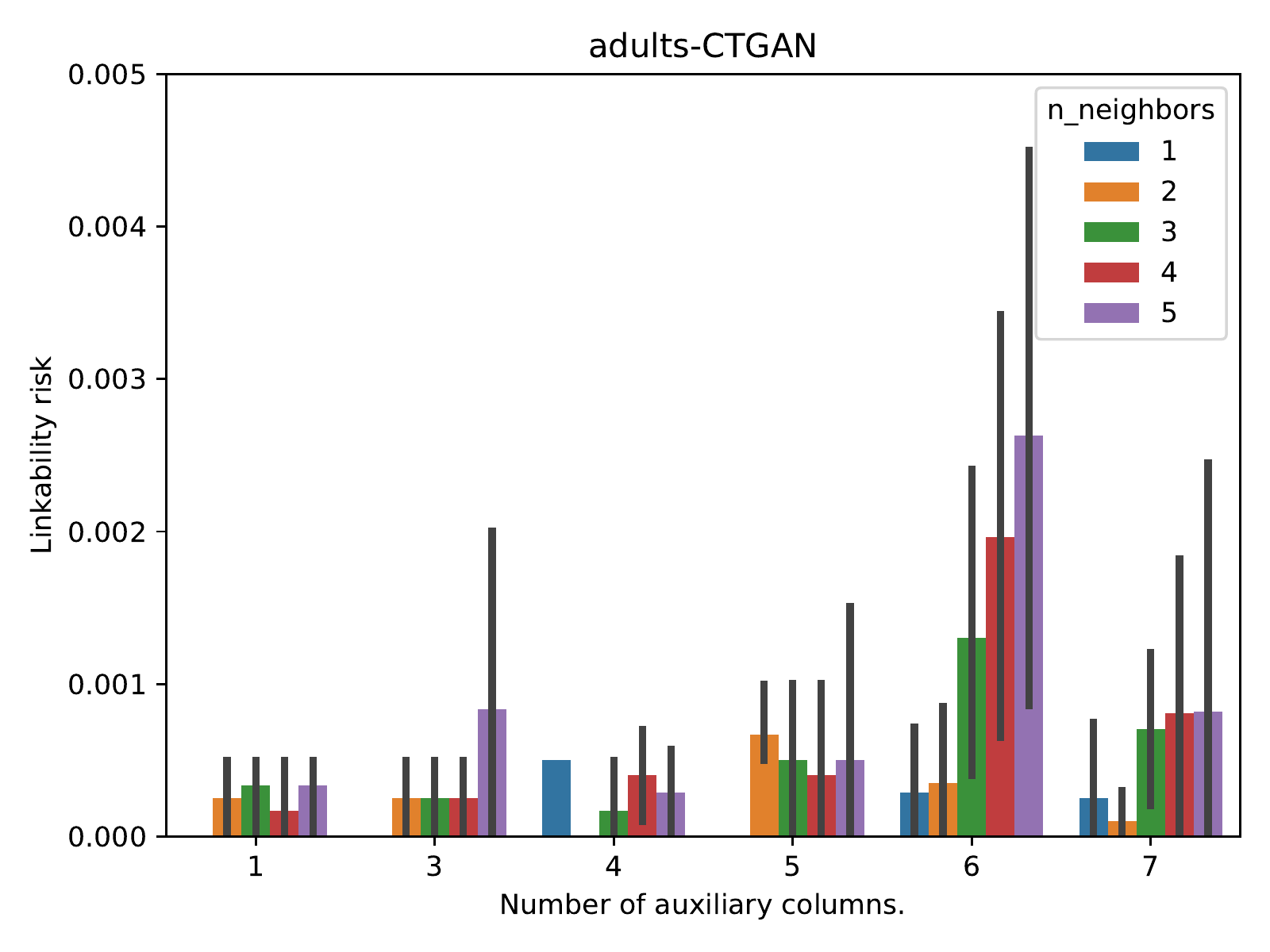}~
\includegraphics[width=0.25\textwidth]{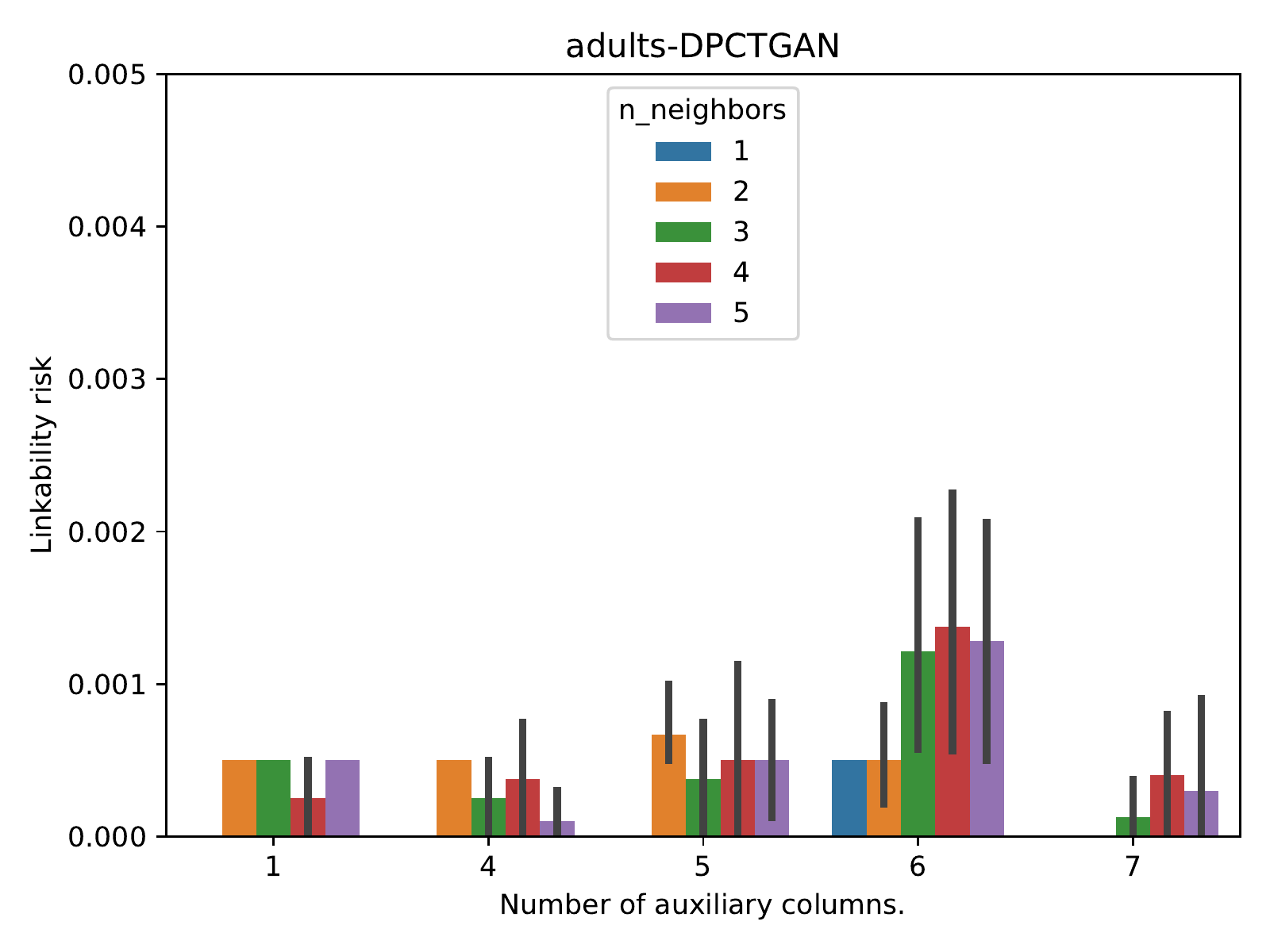}
\caption{Measured linkability risk in the Adult dataset as a function of the amount of auxiliary columns used to mount the attack and and the number $k$ of nearest neighbors considered in the evaluation phase. The two plots refer to synthetic data generated with CTGAN (left) and DPCTGAN (right). For more details, see~\Cref{subsec:linkability}.\label{fig:linkability_num_neighbors}}
\end{figure}

\begin{figure}[tbh]
\centering
\includegraphics[width=0.16\textwidth]{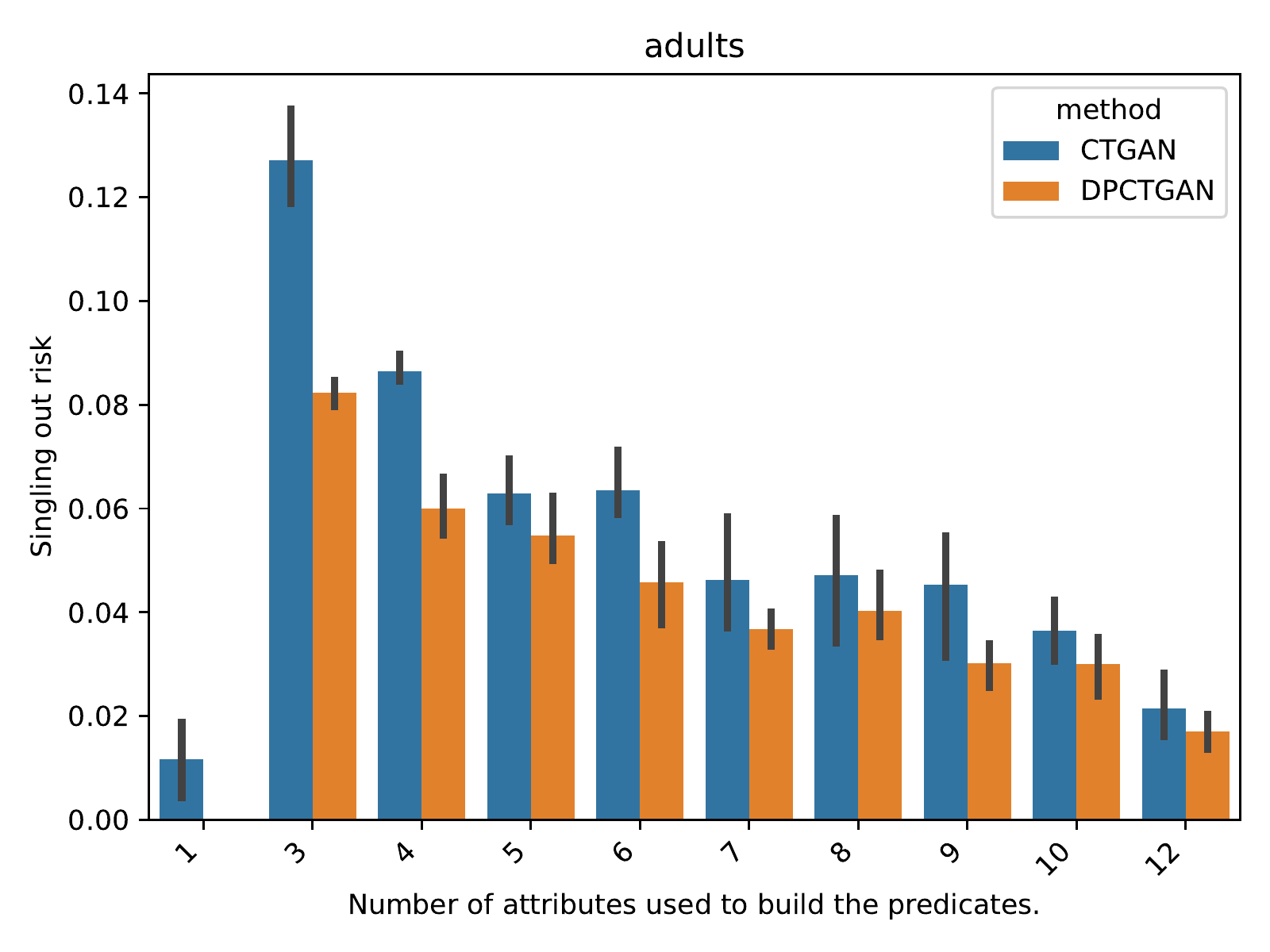}~
\includegraphics[width=0.16\textwidth]{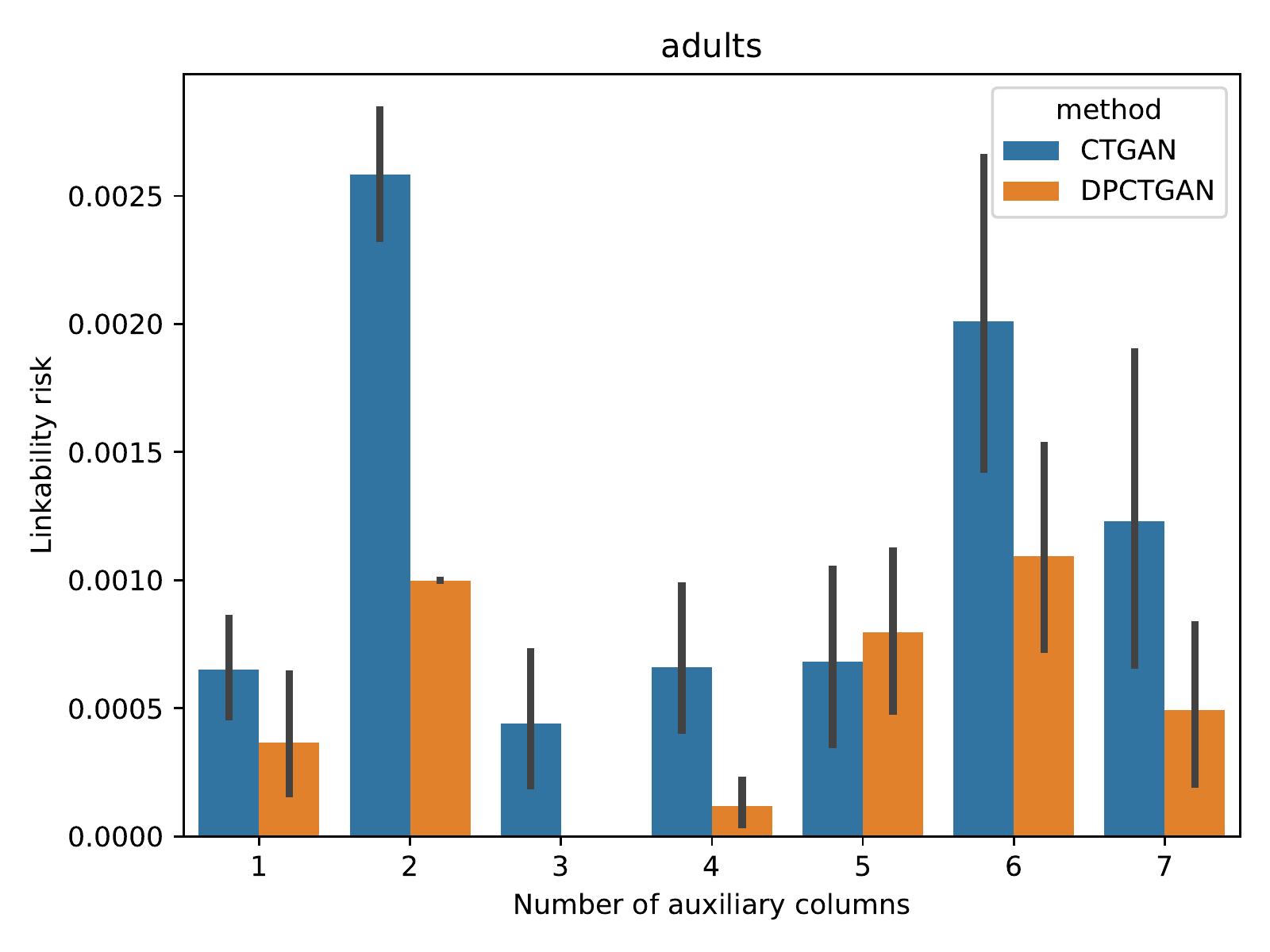}~
\includegraphics[width=0.16\textwidth]{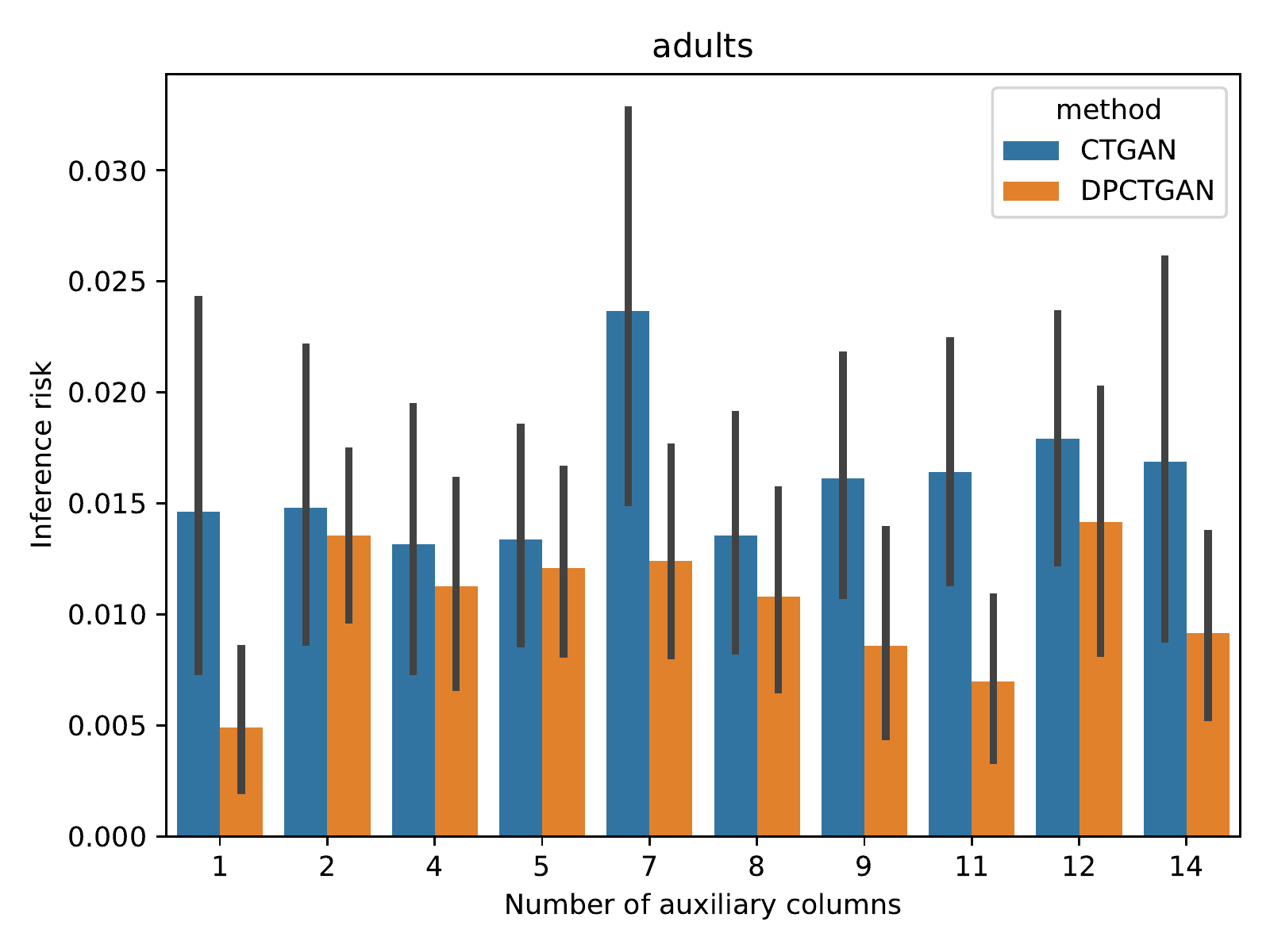}
\caption{Singling out (left), linkability (middle), and inference (right) risks as a function of the amount of auxiliary information (in the case of linkability and inference) or number of attributes used in the predicates (for singling out) for the Adults dataset. See~\Cref{subsubsec:setup} for details.}
\label{fig:risks_vs_aux}
\end{figure}

\end{document}